\documentclass[11pt]{article}
\usepackage{geometry}
\usepackage{nicefrac}
\usepackage{latexsym}
\usepackage{epsf,graphicx,amssymb,stmaryrd}
\usepackage{amsfonts,float,amsmath}
\usepackage{subfigure,epsfig}
\usepackage{hhline,xspace,rotate}
\usepackage{psfrag}
\usepackage{verbatim}

\newfont{\twelvemsb}{msbm10 scaled\magstep1}
\newfont{\eightmsb}{msbm8}
\newfam\msbfam
\textfont\msbfam=\twelvemsb \scriptfont\msbfam=\eightmsb
\catcode`\@=11
\def\Bbb{\ifmmode\let\next\Bbb@\else
\def\next{\errmessage{Use \string\Bbb\space only in math mode}}\fi\next}
\def\Bbb@#1{{\fam\msbfam{{#1}}}}

\renewcommand{\theequation}{\thesection.\arabic{equation}}
\begin{document}

\sloppy
\renewcommand{\thefootnote}{\fnsymbol{footnote}}
\setcounter{page}{1} \vspace{0.1cm}

\def\sm{${\cal{S}} {\cal{M}}\;$}
\def\sms{${\cal{S}} {\cal{M}}$}
\newcommand{\noi}{\noindent}

\newcommand{\ra}{\rightarrow}
\newcommand{\beq}{\begin{equation}}
\newcommand{\eeq}{\end{equation}}
\newcommand{\beqn}{\begin{eqnarray}}
\newcommand{\eeqn}{\end{eqnarray}}
\newcommand{\epem}{e^+ e^-}
\newcommand{\epemt}{$e^+ e^- \;$}
\newcommand{\zzttt}{$ Z Z \ra t \bar t \;$}
\newcommand{\zzttp}{$ Z Z \ra t \bar t $}
\newcommand{\zztt}{$ Z Z \ra t \bar t $ }
\newcommand{\wwttt}{$ W^- W^+ \ra t \bar t \;$}
\newcommand{\wwttp}{$ W^- W^+ \ra t \bar t $}
\newcommand{\wwtt}{$ W^- W^+ \ra t \bar t $}

\begin{titlepage}

\vspace*{0.1cm}\rightline{LAPTH-1207}

\vspace{1mm}
\begin{center}

{\Large{\bf One-loop Electroweak and QCD corrections to vector
boson scattering into top pairs
 and application to the ILC }}

\vspace{.5cm}

N.\ Bouayed ${}^{1,2)}$ and  F.~Boudjema${}^{2)}$

\vspace{4mm}

{\it 1) Universit\'e S\^aad Dahlab de Blida, route de Soum\^aa, B.P.270, 09000 Blida,
Algeria}\\
{\it 2) LAPTH${\;}^\dagger$, Universit\'e de Savoie, CNRS, 9 chemin de Bellevue,  B.P.110,\\ F-74941 Annecy-le-Vieux, France.}

\vspace{10mm} \abstract{ We calculate the electroweak and QCD
corrections to \wwttt and \zzttt. We also consider the interplay
of these corrections with the effect of anomalous interactions
that affect the massive weak bosons and the top. The results at the $VV$ level
fusion are convoluted with the help of the effective vector boson
approximation to give predictions for a high energy \epemt
collider.}

\end{center}

\vspace*{\fill} $^\dagger${\small Laboratoire d'Annecy-le-Vieux de Physique Th\'eorique, UMR 5108.} \normalsize
\end{titlepage}

%%%%%%%%%%%%%%%%%%%%%%%%%%%%%%%%%%%%%%%%%%%%%%%%%%%%%%%%%%%
%%%%%%%%%%%%%%%%%%%%%%%%%%%%%%%%%%%%%%%%%%%%%%%%%%%%%%%%%%%

\section{Introduction}

The fact that the Higgs has not been discovered yet is evidence
that the mechanism behind electroweak symmetry breaking, EWSB,
giving masses not only to the weak vector bosons but also to
fermions remains a mystery. Moreover the implementation of the
Higgs within the Standard Model, \sms, poses a few conceptual
problems. This has triggered an intense activity in model building
in order to solve the hierarchy problem leading to a rich
phenomenology at the upcoming high energy colliders, such as the
Large Hadron Collider (LHC)\cite{LHC} and the International Linear
Collider (ILC) \cite{ILC}.

In practically all the New Physics models which address the
hierarchy problem the top plays a central and crucial role. One
can mention models of dynamical electroweak symmetry breaking in
the technicolour vein where  the concept of top
condensation\cite{top-dynssb} is exploited. In supersymmetry, the
top keeps the the so-called minimal model alive through radiative
corrections to the Higgs while in mSUGRA, for example,  the top
can trigger electroweak symmetry breaking. The phenomenology of
the top in models of extra-dimensions is also special as discussed
very recently\cite{top_kk}. It is no wonder that the top plays
such a central role in probing the sector of EWSB as well as the
flavour problem. After all, its large mass\cite{topmass} is of the
order of the spontaneous electroweak symmetry scale $v/\sqrt{2} =
174 \, GeV$ which, in the \sms, gives it a Yukawa coupling of
order one, incidentally sensibly of the same strength as the QCD
coupling constant at the EWSB scale. This also explains why
radiative corrections involving the top, especially in processes
related to symmetry breaking, may be competitive with  QCD
corrections. It is in this context that we compute in this paper
the electroweak corrections to vector boson fusion into a pair of
top quarks, \zztt and \wwtt. Weak vector boson fusion is the
privileged process revealing spontaneous symmetry breaking through
the longitudinal or Goldstone component of the vector bosons.
Although most analyses have dealt with scattering into vector
bosons, it is clear that fusion into tops is also a probe of
symmetry breaking. Since the effects of the EWSB due to New
Physics  can be subtle, it is important that we know quite
precisely the \sm prediction. This calls for calculations beyond
tree-level. QCD corrections have been performed recently in
\cite{GZ05}. We will include here the electroweak corrections
alongside the QCD corrections. We also compare the loop
corrections effects with  those from new operators parameterising
the EWSB effects in \wwtt. The complete study of vector boson
fusion at the colliders would require a full $2 \ra 4$
calculation. Especially for the inclusion of radiative corrections
this is a most daunting task. Though most of the machinery has
been developed to tackle such complex $2 \ra 4$ processes at loop
level, there have been till now only two such
computations\cite{eeto4fdenner,grace_eetonnhh}. It therefore seems
fit in a first step to revert to some approximation. The effective
vector boson approximation, EVBA\cite{EVBA}, seems suited for such
processes. This approximation has been extensively used in the
context of weak vector fusion. We will conduct our calculation
within this approximation for application to the \epemt linear
collider. The important genuine electroweak corrections should be
captured in the subprocess. We only restrict ourselves to the
linear collider; \wwttt in the context of the LHC suffers from
considerable background\cite{LHC_WWtott} mainly due to the
overwhelming {\em direct} $t \bar t$ production. The Higgs range
we consider in this paper does not go beyond a Higgs mass,
$M_{Higgs}$, of $300 \,$GeV in accordance with the current limit
from precision measurements\cite{LEPEWWG06}. We intend to study
the radiative corrections in the presence of a very heavy Higgs in
a forthcoming publication.

 This paper is organized as follows. In the next section we
 briefly review \zztt and \wwttt and show the importance of the
 longitudinal modes especially at high energies. We then expose
 the EVBA. In section~3 we present the main ingredients that enter
 the calculations at one-loop, the checks we performed as well as the
 tools we exploited. Section~4 gives our results for the
 electroweak and QCD corrections. Section~5 shows  these
 results after convolution with the EVBA structure functions for
 the linear collider. Section~6 introduces the anomalous operators
 and their effects on the processes we study. Section~7 summarises
 our conclusions. An appendix lists all formulae for the helicity
 amplitudes at tree-level, in a compact way, including the
 anomalous operators.

%%%%%%%%%%%%%%%%%%%%%%%%%%%%%%%%%%%%%%%%%%%%%%%%%%%%%%%%%%%
%%%%%%%%%%%%%%%%%%%%%%%%%%%%%%%%%%%%%%%%%%%%%%%%%%%%%%%%%%%

\section{Tree-level overview and dominance of the longitudinal modes}
\setcounter{equation}{0}

%%%%%%%%%%%%%%%%%%%%%%%%%%%%%%%
\subsection{Notation and Born contribution}
%%%%%%%%%%%%%%%%%%%%%%%%%%%%%%%
\begin{figure}[ht]
\begin{center}
  \subfigure[$ZZ \to t\bar{t}$ \label{Born_zztt}]{\epsfig{figure=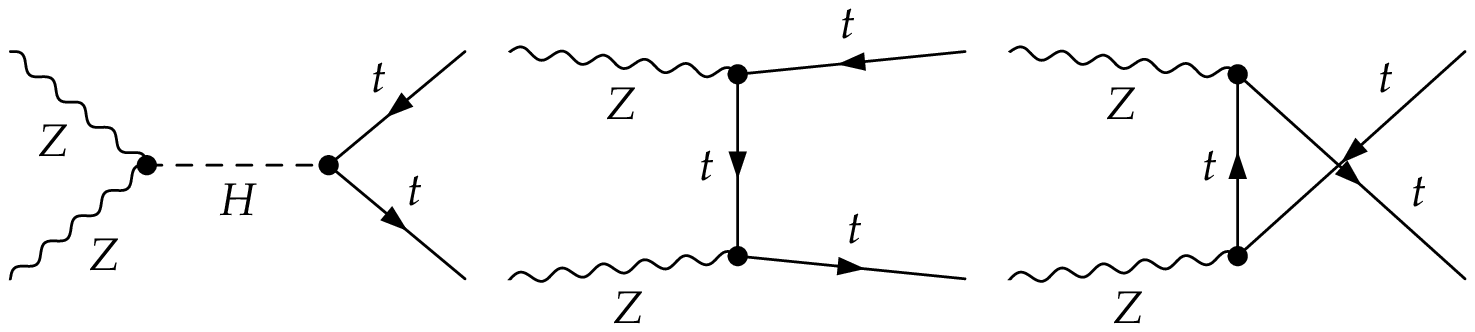,width=0.42\textwidth}}
\hspace{1cm}
  \subfigure[$W^-W^+ \to t\bar{t}$ \label{Born_wwtt} ]{\epsfig{figure=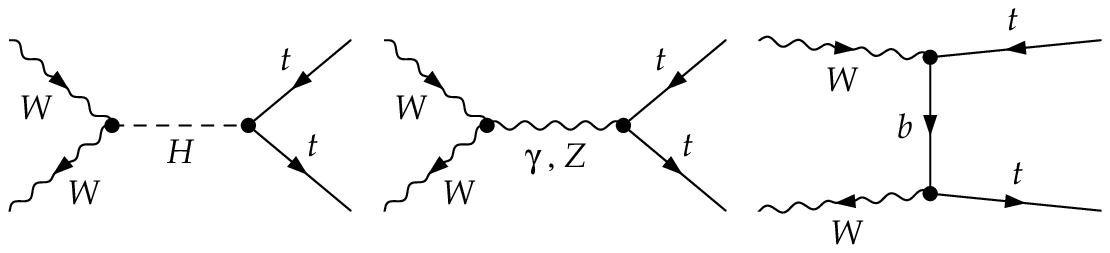,width=0.42\textwidth}}
\end{center}
\caption{{\em Tree level Feynman diagrams for the process a) ${ZZ
\to \bar{t}t}$ and b) ${W^-W^+ \to \bar{t}t}$.} }
\label{one-loop-diagrams}
\end{figure}

The tree level Feynman diagrams for the processes we study,
$Z(k_1,\lambda_1)+Z(k_2,\lambda_2)\to
t(k_3,;\lambda_3)+\bar{t}(k_4,\lambda_4)$ and
$W^-(k_1,\lambda_1)+W^+(k_2,\lambda_2)\to
t(k_3,;\lambda_3)+\bar{t}(k_4,\lambda_4)$ are shown in figure~
\ref{Born_zztt} and \ref{Born_wwtt}. The labelling $k_i$ and
$\lambda_i=\pm,0$ stand for the momentum and the helicity.
Complete helicity amplitudes formulae for \wwttt are given in the
Appendix.

\begin{figure}[H]
\begin{center}\includegraphics[width=0.7\textwidth,
  keepaspectratio]{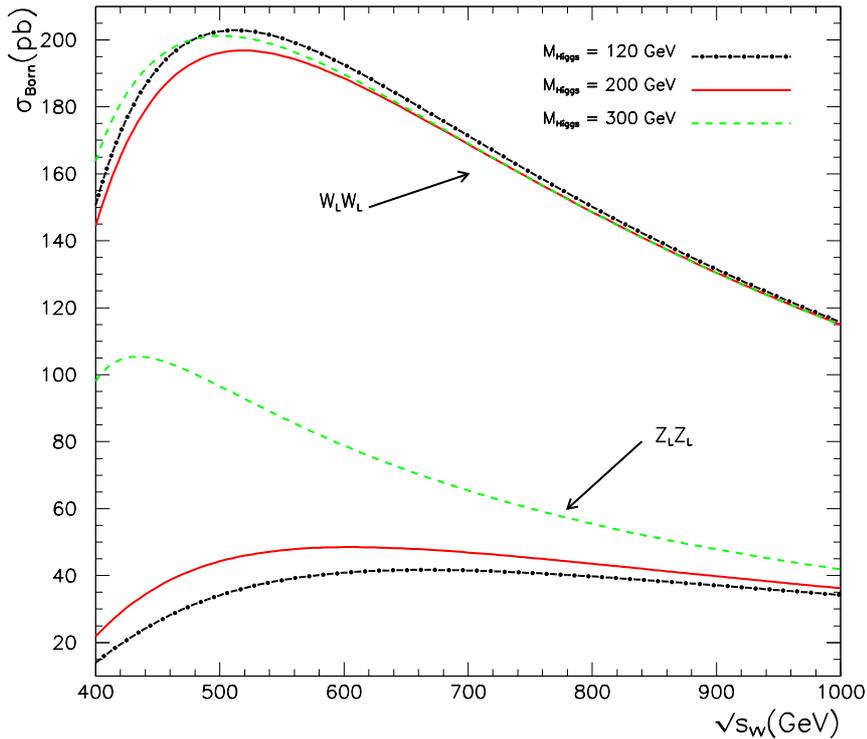}\end{center}
\caption{\label{Born-vvtt-LL} {\em Total Born cross section for
the processes ${W_L^-W_L^+ \to \bar{t}t}$ and ${Z_LZ_L \to
\bar{t}t}$ for  $M_{Higgs}= 120, 200, 300 \,GeV$ and  $p_T^{t,\bar
t}>10$\,GeV.}}
\end{figure}

We are mostly interested in the longitudinal modes of the vector
bosons, with $\lambda_1=\lambda_2=0\equiv L$. Total cross sections
for this mode are shown in Figs~\ref {Born-vvtt-LL}.
$Z_L Z_L$ is more sensitive to the Higgs mass dependence
especially for low centre of mass energies. The $W$ mode provides
though more yield. Note also, that for the range of Higgs masses
we are studying the cross sections after reaching a maximum around
threshold, drop as the energy increases. This should be taken into
account when considering the effect of New Physics (NP) that tends
to increase the cross section as the energy increases. When
convoluting with the ``structure function" of the EVBA (see next),
the lower end of the cross section might pollute the NP effects.
One should therefore aim at imposing a cut on the invariant mass
of the system. We will always take a cut on the invariant mass of
the $t \bar t$, $m_{t \bar t}> 400$GeV as is done in\cite{GZ05}.
This cut will help improve the EVBA. We will also impose a cut on
the transverse momentum of the top, $p_T^{t,\bar t}>10$GeV. We
will do this not only after convolution on the structure function,
{\it i.e.} at the \epemt level but also at the $VV\ra t \bar t$
level. The figures we show are with these cuts. The $p_T^{t,\bar
t}$ cut helps reduce some QCD background and fake $p_T$ in $t \bar
t \gamma$, see\cite{Barklow_wwtt,Alcaraz_wwtt,LariosTait}.

\begin{figure}[h]
\begin{center}
\mbox{\includegraphics[width=0.49\textwidth,height=0.60\textwidth]{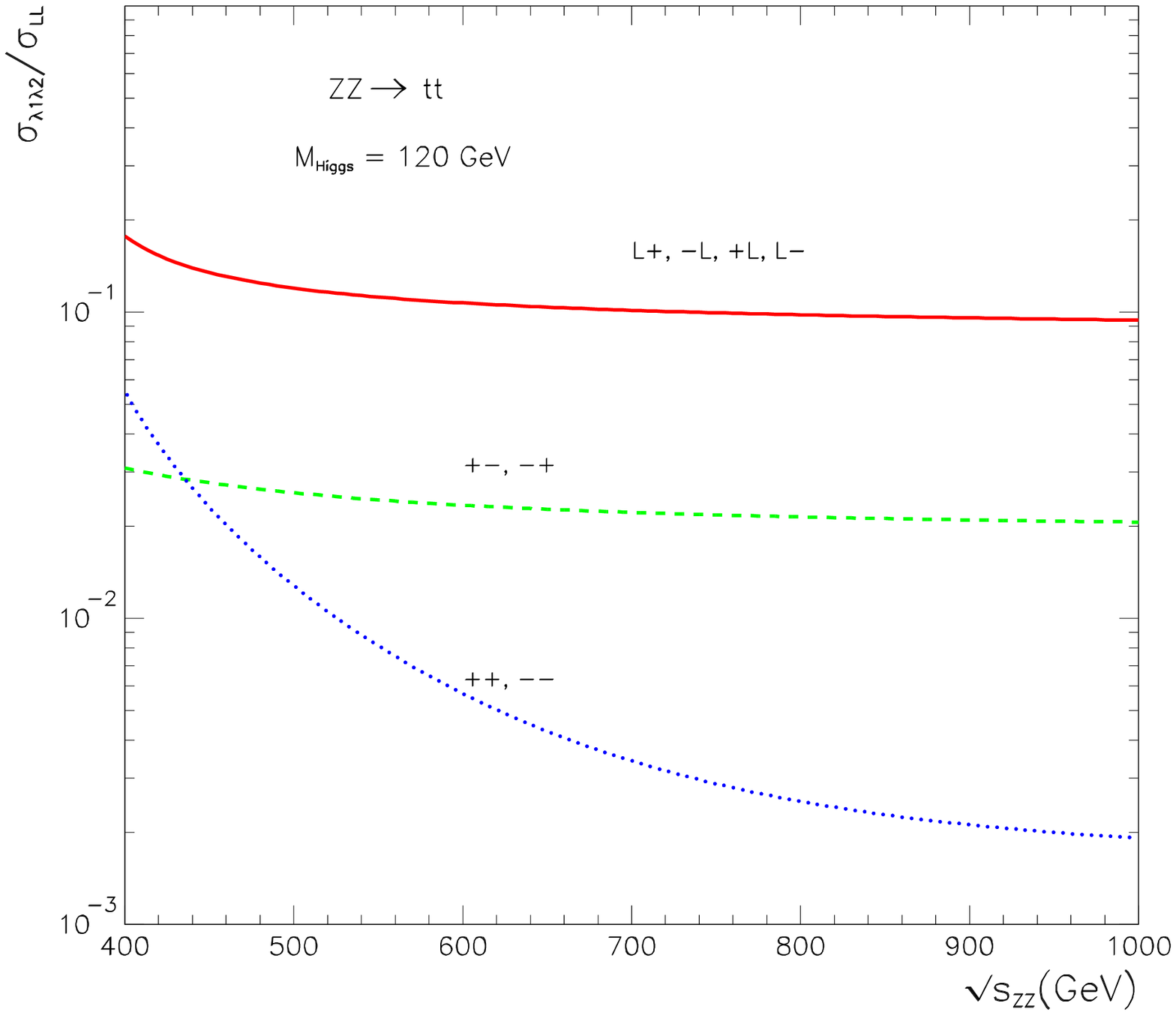}
\includegraphics[width=0.49\textwidth,height=0.60\textwidth]{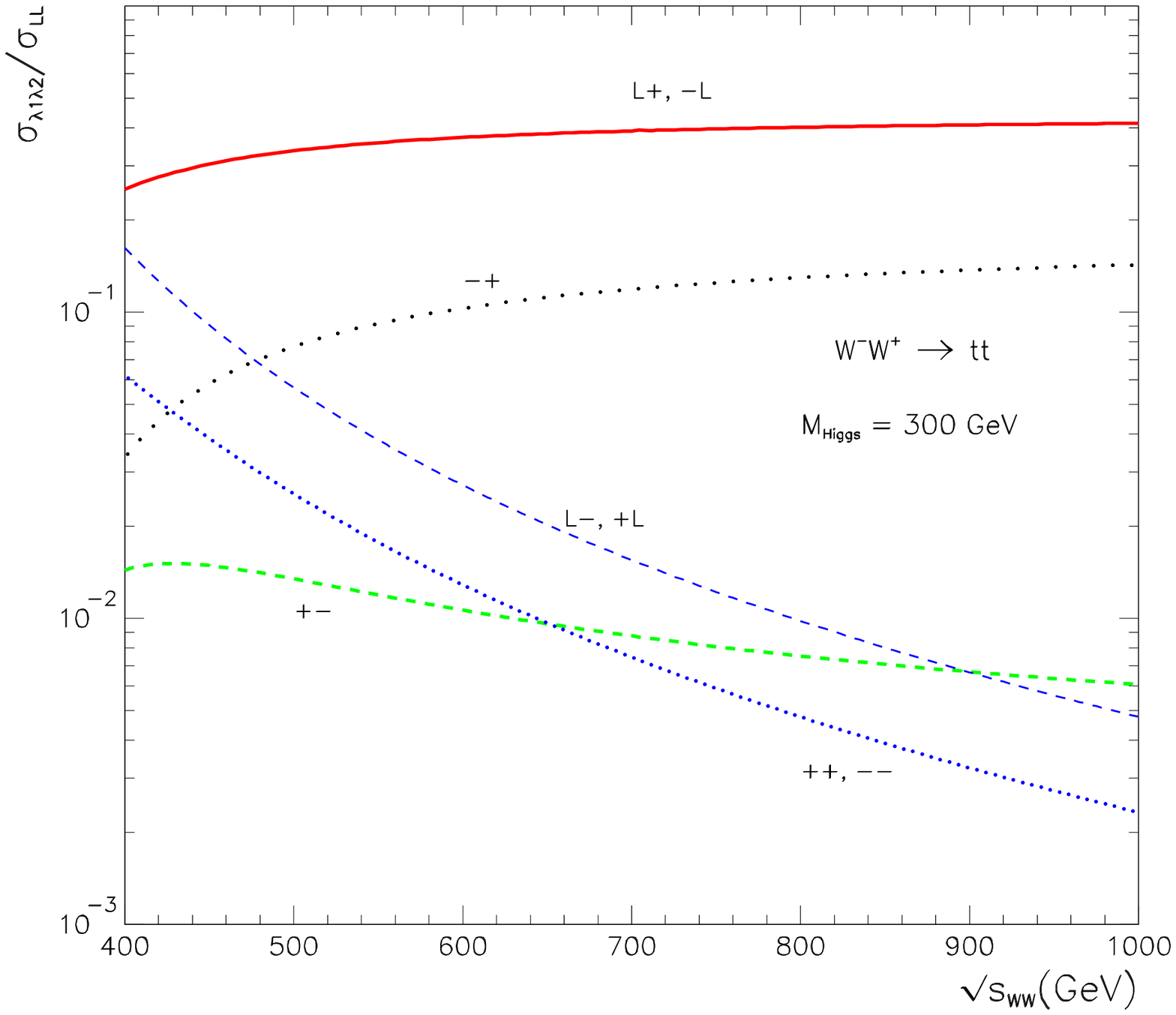}}
\caption{{\em Relative rates for the Born cross section for
different polarisations of the incoming vector bosons with respect
to the case where the incoming bosons are both longitudinally
polarised. For $WW$, $\lambda_i \lambda_j$ labelling of the
helicities refers to $W^-(\lambda_i)
W^+(\lambda_j)$.}}\label{k-polarization-tree}
\end{center}
\end{figure}
Another reason for aiming at a higher cut on the invariant $VV$
(or similarly $t \bar t$) system can be gleaned in
Fig.~\ref{k-polarization-tree}. First $V_L V_L$ scattering
dominates for all energies. This said, at the lowest centre of
mass energies, some transverse contributions might not be so
negligible as compared to $V_L V_L$ fusion, though they drop
rather rapidly as the energy increases. Note however that the
longitudinal/transverse ($V_L V_T$) mode in the \sm might bring
some contamination at all energies, which could worsen the EVBA
approximation.

\subsection{Born angular distributions}
\begin{figure}[H]
\begin{center}
\mbox{\includegraphics[width=0.45\textwidth,height=0.45\textwidth]{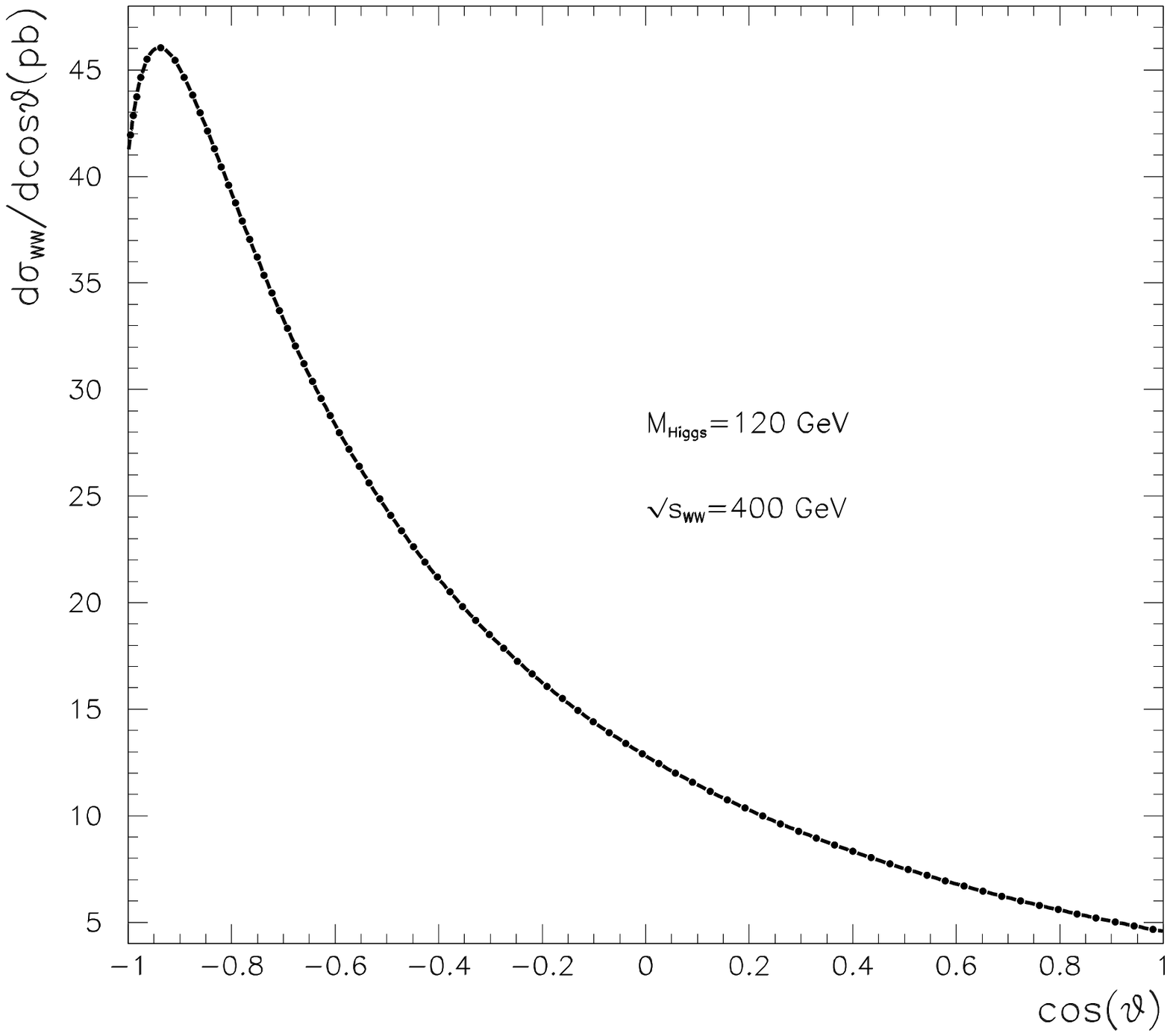}
\includegraphics[width=0.45\textwidth,height=0.45\textwidth]{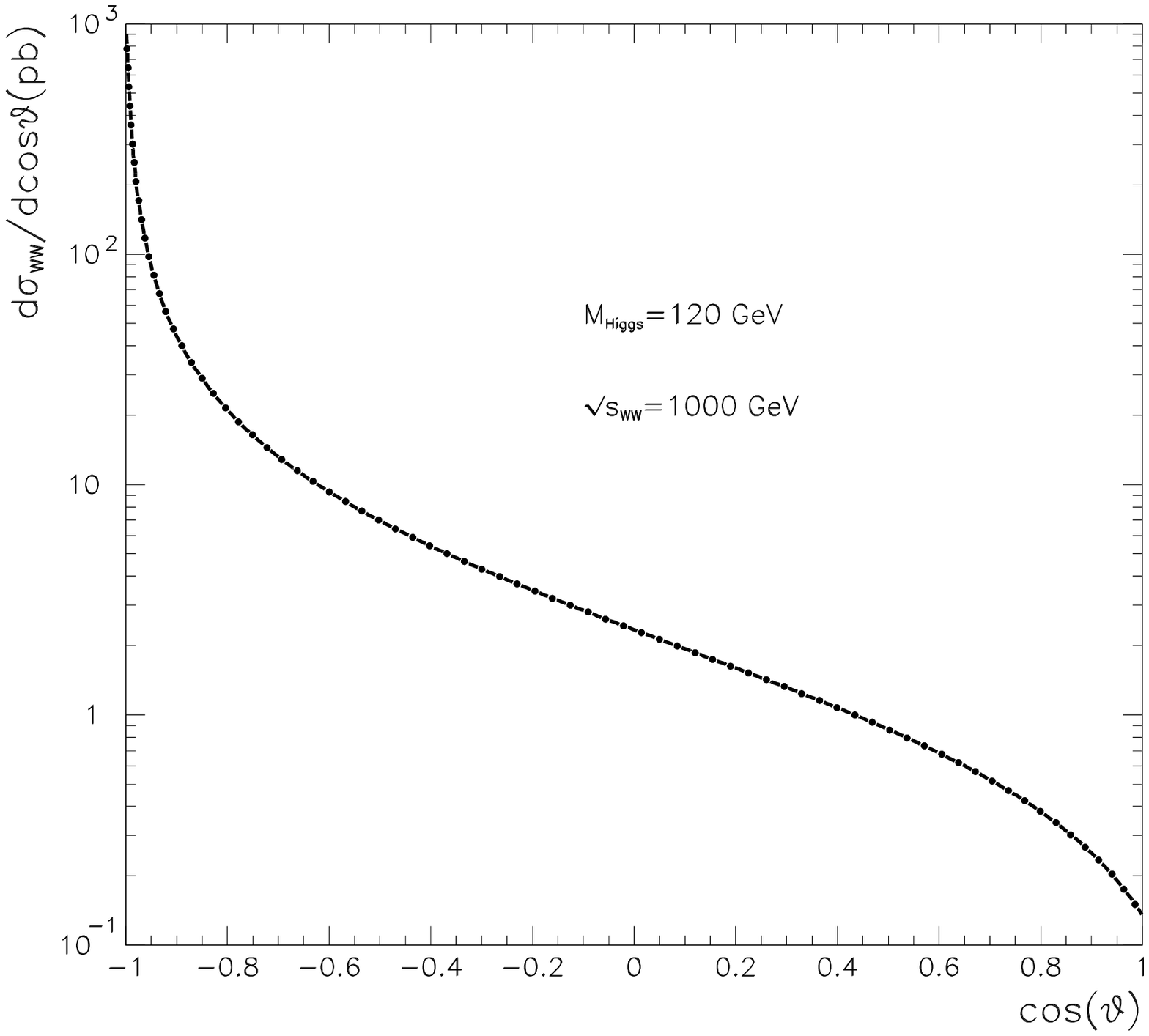}}
\caption{{\em Angular distribution for the
unpolarised \wwtt \, cross section for $M_{{\rm Higgs}}=120$GeV at
$\sqrt{s_{WW}}=400$GeV and $1$TeV. $\theta$ is the angle between
the $W^-$ and the $t$. The $p_T^{t,\bar t}>10$\,GeV corresponds to a
too small $\theta_{\rm cut}$ that can not be represented on the
plot.}}\label{diffcross-tree}
\end{center}
\end{figure}

Radiative corrections and the New physics can not only change the
total yield but can also, and most dramatically, distort the shape
of kinematical distributions. We show in Fig.~\ref{diffcross-tree}
the angular distribution in \wwtt \,for $\sqrt{s_{WW}}=400$\,GeV,
not too far from threshold and at high energy
$\sqrt{s_{WW}}=1$\,TeV. We choose to show \wwtt \, because of the
chiral structure of the $W$ coupling and the small mass of the
bottom (exchanged in the $t$-channel). This leads, at high energy,
to an overwhelming production of top in the backward region, with
respect to $W^-$. In $ZZ$, the distribution are of course
symmetric and  much less peaked. A cut on the forward backward
region will also help bring out the signal of New Physics which
occurs at higher $p_T$.

%%%%%%%%%%%%%%%%%%%%%%%%%%%%%%%
\subsection{Weak bosons effective distribution functions}
%%%%%%%%%%%%%%%%%%%%%%%%%%%%%%%
To turn the subprocess \zzttt and \wwtt \, into $e^+e^- \to
t\bar{t}e^+e^- $  and $e^+e^- \to t\bar{t}\nu_e\bar{\nu}_e$
respectively, we revert to the EVBA approximation\cite{EVBA}. The
approximation  introduces the notion of the $W/Z$ content of the
electron (and positron). At high energies, we take the structure
functions of the weak vector bosons inside a fermion as given by
\cite{Vincent96,testEVA} :

\begin{equation}
f_{fermion/V,\lambda}(x) = \frac{1}{16 \pi x} \Big [ (g_V -
\lambda g_A)^2 + (1-x)^2(g_V + \lambda g_A)^2 \Big ] \Big [\ln
\frac{P_{T,max}^2}{M_V^2} - 1 \Big ] ,\quad \lambda = \pm 1
\end{equation}

\begin{equation}
f_{fermion/V,\lambda}(x) = \frac{g_V^2 + g_A^2}{4 \pi
^2}\frac{1-x}{x}, \quad \lambda = 0,
\end{equation}

\noindent $x$ is the momentum fraction transferred to the vector boson by
the initial fermion. $g_V$ and $g_A$ are the vector and
axial-vector couplings of the weak vector bosons to fermions. The
transverse distribution is subject to more uncertainty due to the
logarithmic factor. In our case, we take $P_{T,max}$ the
maximum transverse momentum of the vector bosons allowed by the
kinematics. In the collision of two elementary fermions $f_1$ and
$f_2$ of respective charge and isospin assignement $(Q_{f_1},
T_{f_1}^{(3)})$ and $(Q_{f_2}, T_{f_2}^{(3)}) $, the effective
luminosity of the longitudinal vector bosons is given by:
\begin{equation}
{\frac{d \mathcal{L}}{d {\tau}} \Big{|}}_{f_1f_2/V_LV_L}(\tau) = \int_\tau^1 \frac{dx}{x} f_{f_1/V_L}(x)
 f_{f_2/V_L}(\frac{\tau}{x}).\\ \label{lumVL-leptons}
\end{equation}
This gives for example for $W_L W_L$\cite {KS96}:
\begin{equation}
{\frac{d \mathcal{L}}{d {\tau}} \Big{|}}_{f_1f_2/W_LW_L}(\tau)
={{\Big [}\frac{\alpha} {4\pi \scriptstyle{S_W^2}}{\Big ]}}^2
\frac{1}{\tau}{\Big (}(1+\tau)\ln\frac{1}{\tau}-2(1-\tau){\Big )}.
\label{lumffWWLL}
\end{equation}

The total cross section for the processes $(e^+e^- \,
\stackrel{\scriptscriptstyle{W_L W_L}}{\longrightarrow} \,t\bar{t}
+ \nu_e \bar{\nu}_e)$ is then given by:

\begin{equation}
%\begin{split}
\hat{\sigma}_{(e^+e^- \, \stackrel{\scriptscriptstyle{W_L
W_L}}{\longrightarrow} \,t\bar{t})}(s_{ee}) = \int_{\tau_{min}}^1
d\tau {\frac{d \mathcal{L}}{d {\tau}}\Big{|}}_{e^+e^-/W_L
W_L}(\tau)
 \tilde{\sigma}_{W_L W_L \to t\bar{t}}(s_{VV}= \tau\, s_{ee} ), \;
 \quad  \tau_{min} = \frac{4 M_{top}^2}{s_{ee}}.
%\end{split} \label{hatilccross}
\end{equation}
Formulae for the $ZZ$ case can be derived in a similar manner.

\begin{figure}[H]
\begin{center}
  \subfigure[$e^+e^- \to t\bar{t}e^+e^-$ through ${Z_LZ_L \to t\bar{t}}$ \label{sig-ilc-zz} ]
  {\epsfig{figure=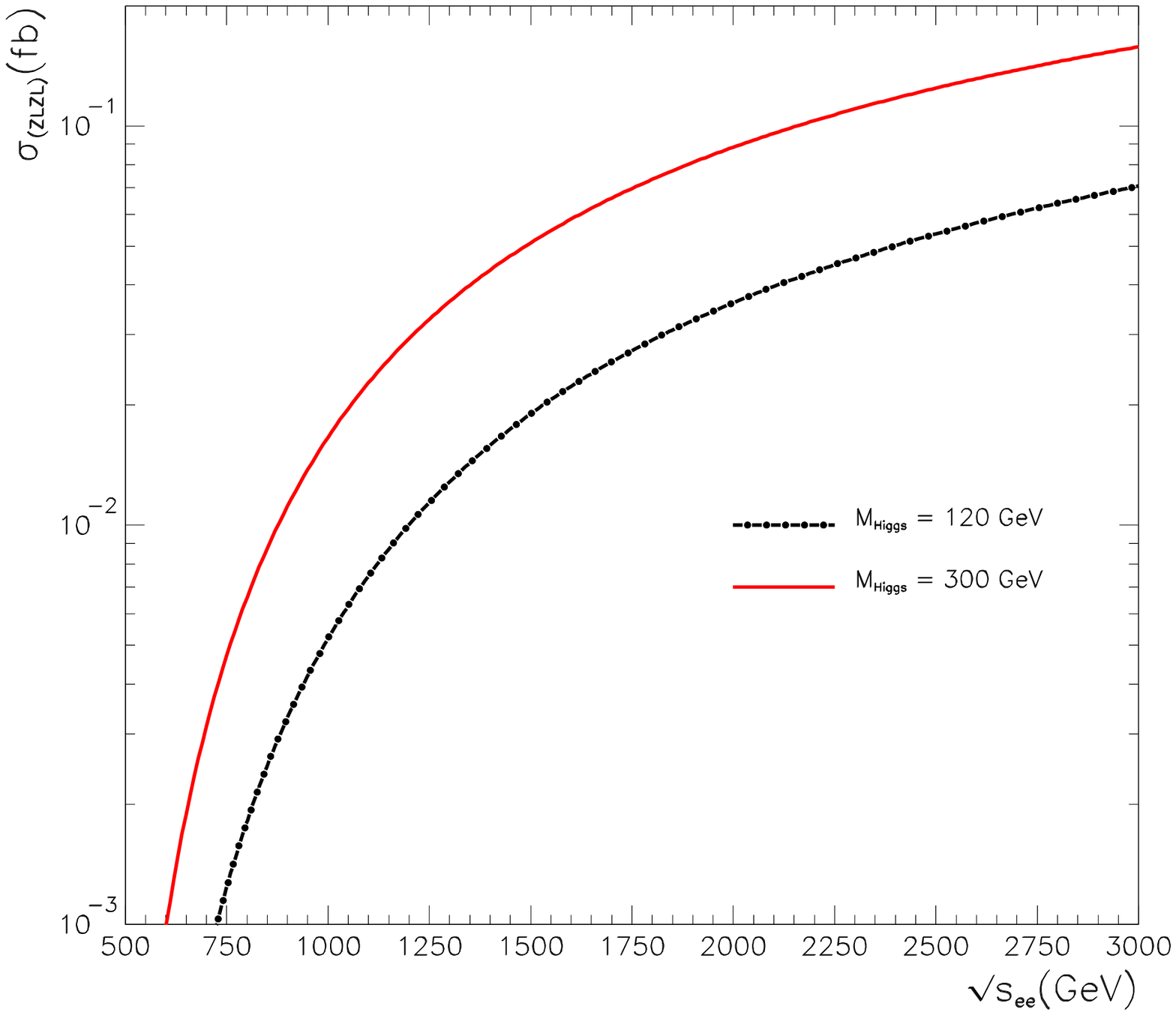,width=8.6cm}}
  \subfigure[$e^+e^- \to t\bar{t}\nu_e\bar{\nu}_e$ through ${W_LW_L \to t\bar{t}}$\label{sig-ilc-ww} ]
  {\epsfig{figure=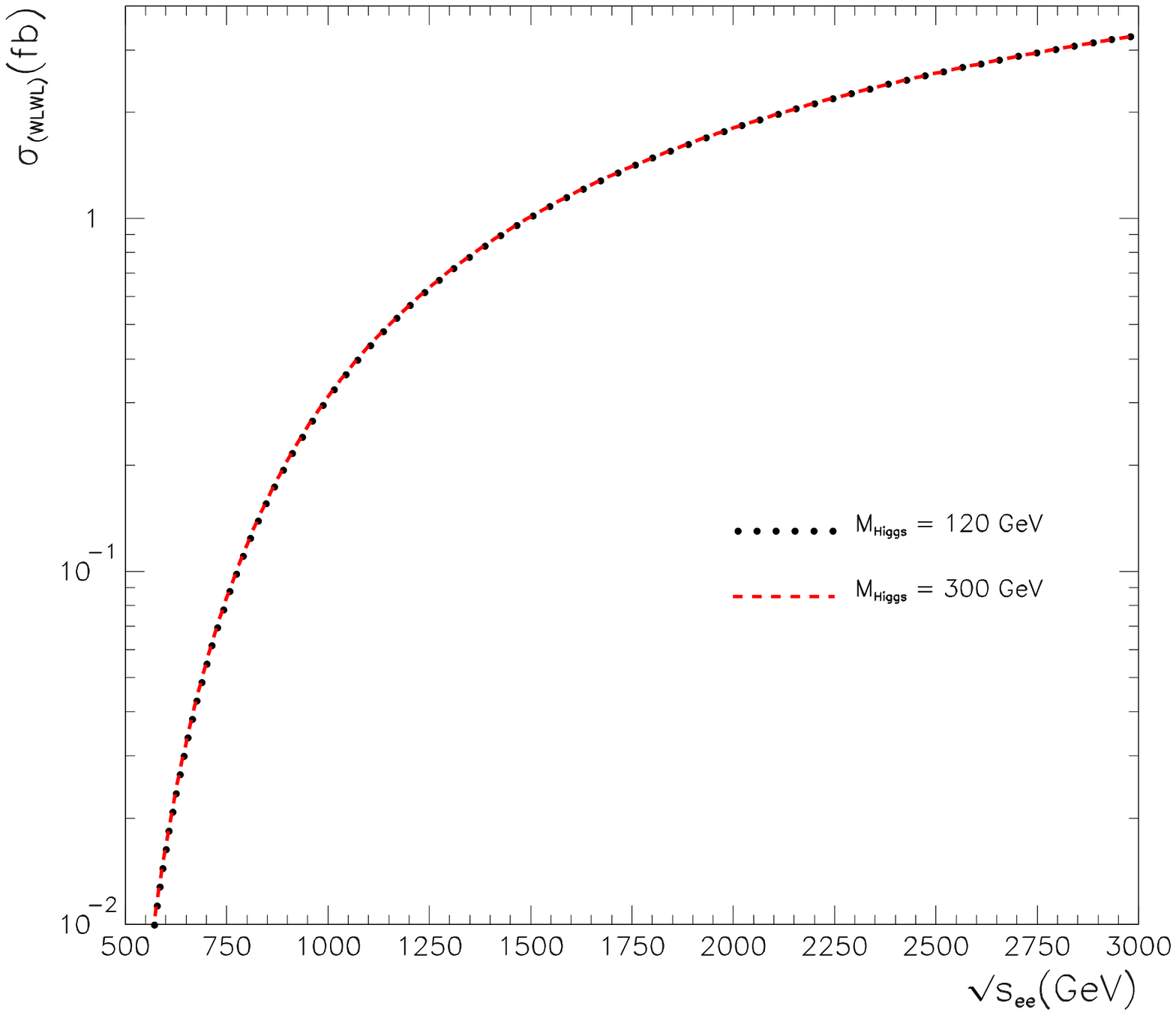,width=8.6cm}}
\end{center}
\caption{ {\em Born total cross sections for (a) $e^+e^- \to
t\bar{t}e^+e^-$ through ${Z_LZ_L \to t\bar{t}}$ and (b) $e^+e^-
\to t\bar{t}\nu\bar{\nu}$ process through ${W_L^- W_L^+ \to
t\bar{t}}$}.} \label{sig-ilc-zz-ww}
\end{figure}

After convolution, the \epemt cross section to $t \bar t$ through
vector boson fusion is reduced much and will most probably be
exploited only in a second stage linear collider with energy in
excess of $1$TeV, see Fig.~\ref{sig-ilc-zz-ww}. Note that at the
level of \epemt, the litte Higgs mass dependence there is in the
$WW$ channel ($M_{Higgs}=120$GeV and $ M_{Higgs}=300$GeV)  is
washed out but not in the $ZZ$ channel, though the latter cross
section is now, at least, an order of magnitude smaller than in
$WW$ suffering from the smaller couplings of the $Z$ to the electrons.

One should keep in mind that the effective EVBA works best at high
energy (compared to the $W$ mass for instance).
Reference\cite{Alcaraz_wwtt} made a comparison of the EVBA with a
full SM calculation (at tree-level) using {\tt
CompHEP}\cite{CompHep}. It was found that the EVBA is a good
approximation for $\sqrt{s_{ee}}=1.5$TeV and above.
%%%%%%%%%%%%%%%%%%%%%%%%%%%%%%%
\section{Electroweak and QCD corrections:\\ Set up and details of
the calculation}
%%%%%%%%%%%%%%%%%%%%%%%%%%%%%%%
\subsection{General structure}
The calculation of the complete one-loop electroweak and QCD
corrections to \wwttt and \zztt  is performed with the help of two
automatic codes for loop corrections, {\tt SloopS}\cite{SloopS}
and {\tt FormCalc/LoopTools}\cite{FormCalc}. Although {\tt SloopS}
uses many modules of \cite{FormCalc}, the model files and the
generation of the Feynman rules is generated automatically with
the help of {\tt LanHep}\cite{lanhep}. Moreover, some loop
integration routines have been improved. The strength of {\tt
SloopS} is the ability to perform very powerful gauge parameter
independence checks as we will see later, a feature that is found
in {\tt GRACE-loop}\cite{Grace-loop}. {\tt SloopS} has been
developed for supersymmetry but has also a \sm module. The results
of {\tt SloopS}, {\tt FormCalc/LoopTools} and {\tt GRACE-loop}
have been checked for a variety of electroweak processes. All
codes adopt the on-shell renormalisation scheme according
to\cite{Grace-loop,eennhletter,kyotorc}. In particular we take as
input, the masses of the particles in the model and the
electromagnetic coupling as defined in the Thomson limit. For the
process at hand, this means $\alpha$ the electromagnetic coupling,
the masses of the $W,M_W$, the $Z, M_Z$ and the top $m_t$.

%%%%%%
\begin{figure}[H]
\begin{center}
  \subfigure[$ZZ \to t\bar{t}$ \label{une-boucle-zztt}]{\epsfig{figure=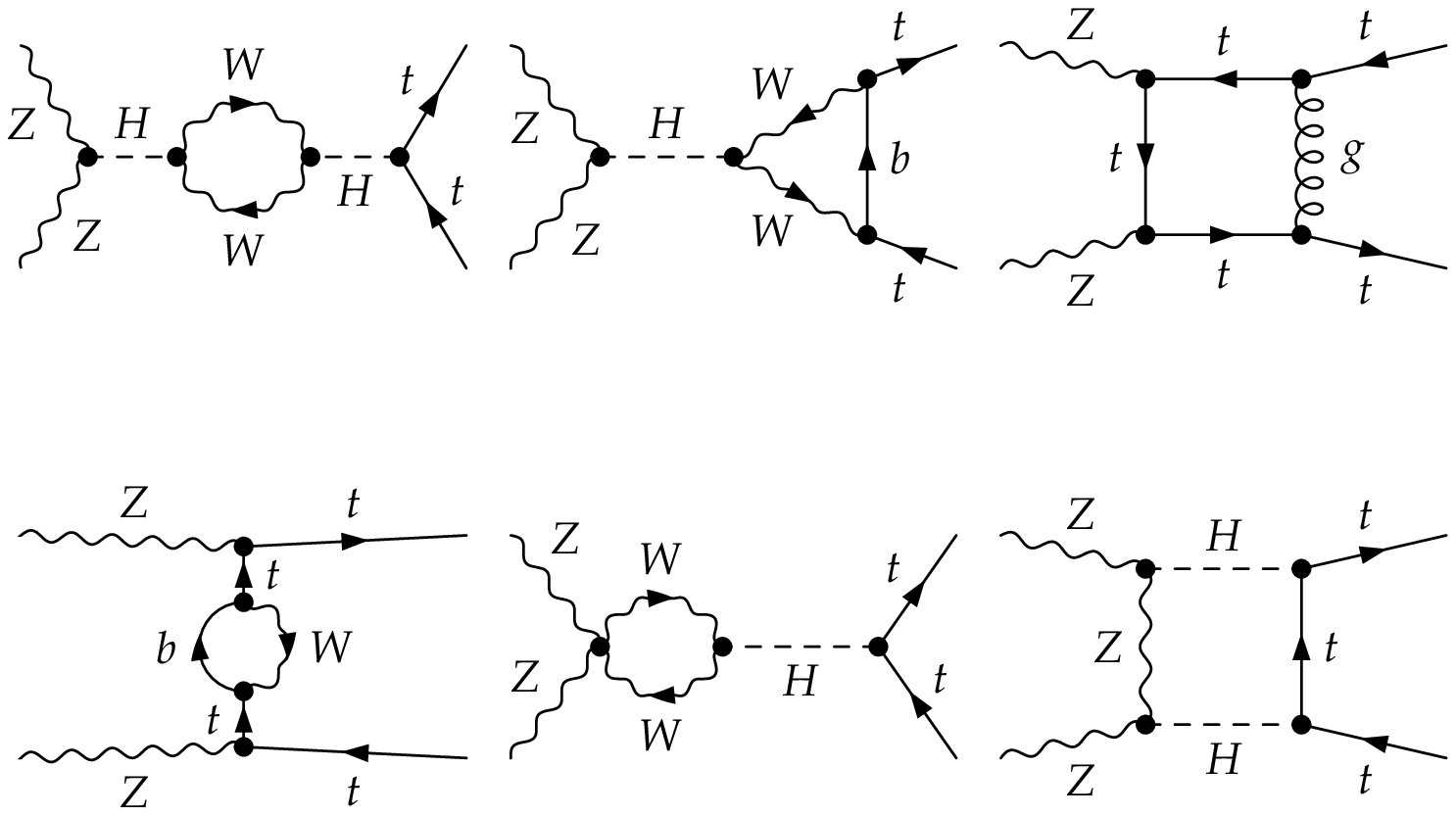,width=7.cm}}
\hspace{1cm}
  \subfigure[$W^-W^+ \to t\bar{t}$ \label{une-boucle-wwtt} ]{\epsfig{figure=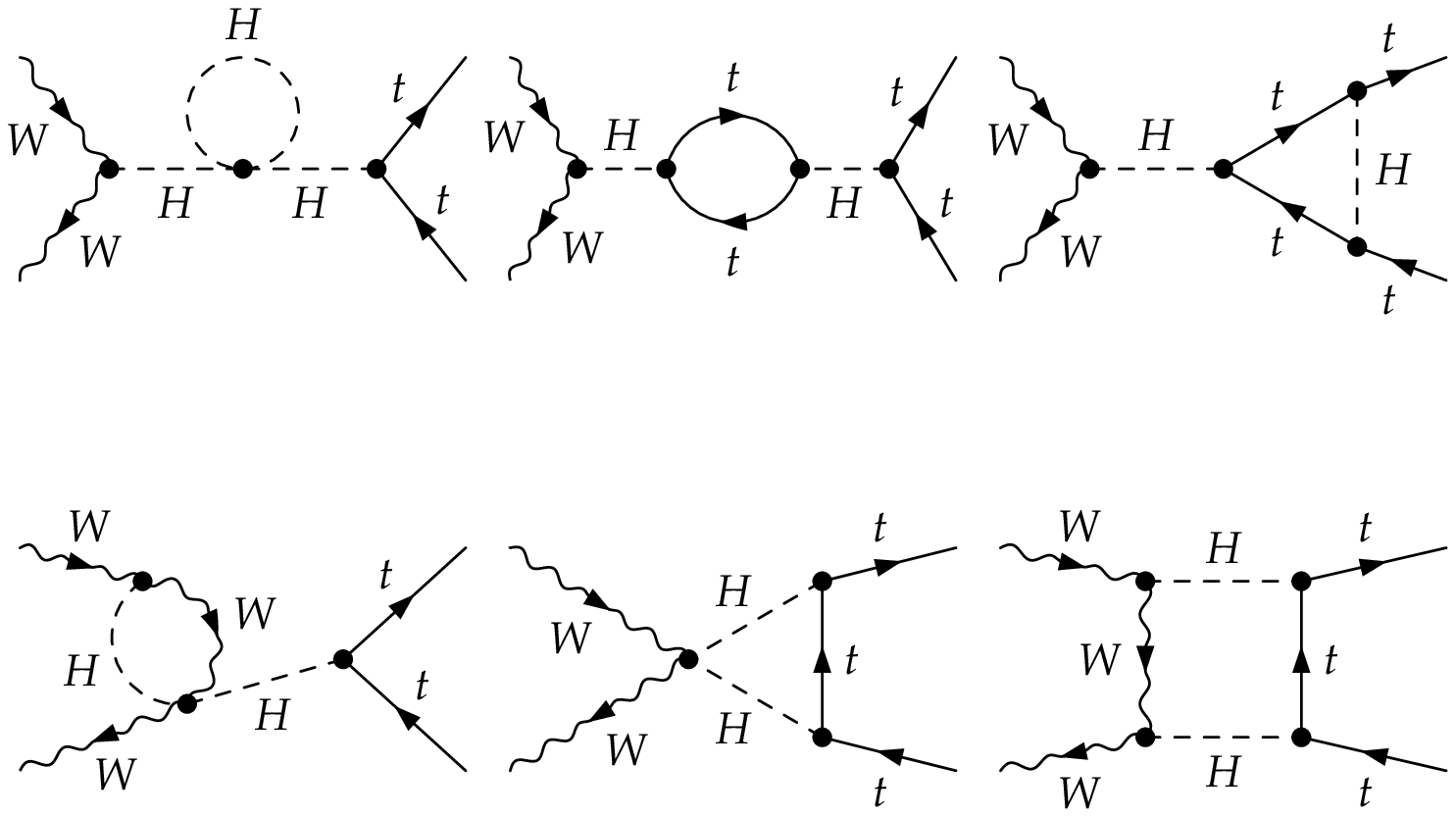,width=7.cm}}
\end{center}
\caption{{\em Some one loop Feynman diagrams for the processes a)
${ZZ \to \bar{t}t}$ and b) ${W^-W^+ \to \bar{t}t}$. }}
\label{one-loop-diagrams}
\end{figure}
\begin{figure}[H]
\begin{center}
  \subfigure[$ZZ \to t\bar{t}g$ \label{Born_zzttg}]{\epsfig{figure=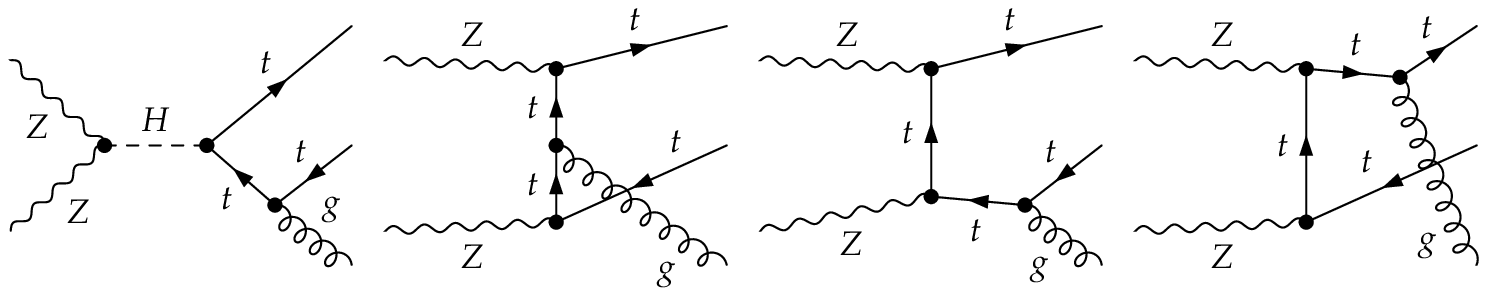,width=7.cm}}
\hspace{1cm}
  \subfigure[$W^-W^+ \to t\bar{t}\gamma$ \label{Born_wwtta} ]{\epsfig{figure=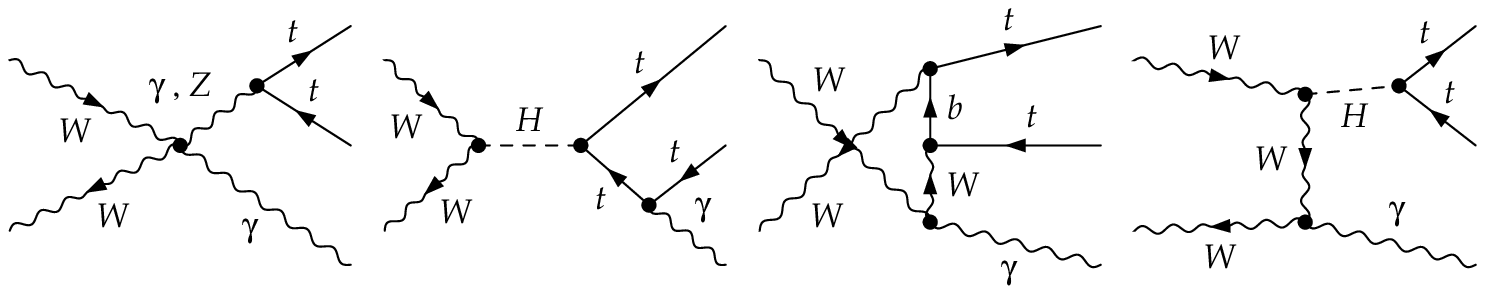,width=7.cm}}
\end{center}
\caption{{\em A selection of the final state radiation diagrams
for the process a) ${ZZ \to t\bar{t}g}$ and b) ${W^-W^+ \to
t\bar{t}\gamma}$. }} \label{born-vvtta}
\end{figure}

The one-loop amplitudes consist of the virtual corrections
$\mathcal{M}_{1loop}^{EW+QCD}$ and the counterterm contributions
$\mathcal{M}_{CT}$. Fig.~\ref{one-loop-diagrams} shows a selection
of the virtual corrections. There are about $500$ diagrams for the
process $W^-W^+ \to t\bar{t}$ and $400$ for the process $ZZ \to
t\bar{t}$. Though $\mathcal{M}_{1loop}^{EW+QCD}+ \mathcal{M}_{CT}$
should be ultraviolet finite, photon and gluon virtual exchange
leads to infrared divergences. These are regulated by a small
photon or gluon mass. For the latter the procedure is admissible
as we are essentially dealing with an Abelian process where the
triple gluon vertex for example does not show up. The photon and
gluon mass regulator contribution  contained in the virtual
correction should cancel exactly against the one present in the
photon and gluon final state radiation. A selection of diagrams in
the latter category is listed in Fig.~\ref{born-vvtta}.
The photonic (gluonic) contribution is in fact split into a soft
part, where the photon (gluon) energy is less than some small
cut-off $k_c$, $\mathcal{M}_{\gamma,g}^{soft}(E_{\gamma,g}<k_c)$
and a hard part with
$\mathcal{M}_{\gamma,g}^{hard}(E_{\gamma,g}>k_c)$. The former
requires a photon/gluon mass regulator. We use the usual universal
factorised form with a simple rescaling for the case of the gluon
correction.  To summarise, the  {\em full} one-loop electroweak
correction is the result of all these contributions:

\begin{equation}
\label{Eq:total_corr}
|\mathcal{M}_1|^{2,EW}  =\overbrace{
2\mathcal{R}e\mathcal{M}_{Born}(\mathcal{M}_{1loop}^{EW}  +
\mathcal{M}_{CT})^* +
|\mathcal{M}_{\gamma}^{soft}|^2}^{|\mathcal{M}_1|^{2,s+v}} +
|\mathcal{M}_{\gamma}^{hard}|^2
\end{equation}

The QCD correction is defined along the same line. The percentage
change, for the {\em total} electroweak correction including hard
photon radiation will be defined as $\delta^{EW}$. For the QCD
correction we will define this as $\delta^{QCD}$.
%%%%%%%%%%%%%%%
\subsection{Checks on the calculation}
%%%%%%%%%%%%%%%

\noi {\it {\bf i)}}  We first check the ultraviolet (UV) finiteness of
the results. This test applies to the whole set  of the virtual
one-loop diagrams. The ultraviolet finiteness test is performed by
varying the dimensional regularisation parameter
$C_{UV}=1/\varepsilon -\gamma_E+\log 4\pi$, $n=4-2 \varepsilon$ is
the dimensionality of space-time. We vary $C_{UV}$ by seven orders
of magnitude with no change in the result. We content ourselves
here and all other tests with double precision. The $C_{UV}$ parameter
could then be set to $0$ in further computations.

\noi {\it {\bf ii)}} The test on the infrared (IR) finiteness is
performed by including both the loop and the soft bremsstrahlung
contributions and checking that there is no dependence on the
fictitious photon mass $\lambda_{\gamma}$ or gluon mass $\lambda_g$.

\noi {\it {\bf iii)}} A crucial test concerns the gauge parameter
independence of the results. Gauge parameter independence of the
result is performed through a set of five gauge fixing parameters.
For the latter a generalised non-linear gauge fixing
condition\cite{Grace-loop,nlg-generalised} has been chosen,
\beqn
\label{fullnonlineargauge} {{\cal L}}_{GF}&=&-\frac{1}{\xi_W}
|(\partial_\mu\;-\;i e \tilde{\alpha} A_\mu\;-\;ig c_W
\tilde{\beta} Z_\mu) W^{\mu +} + \xi_W \frac{g}{2}(v
+\tilde{\delta} H +i \tilde{\kappa} \chi_3)\chi^{+}|^{2} \nonumber \\
& &\;-\frac{1}{2 \xi_Z} (\partial.Z + \xi_Z \frac{g}{ 2 c_W}
(v+\tilde\varepsilon H) \chi_3)^2 \;-\frac{1}{2 \xi_A} (\partial.A
)^2,  \quad c_W=M_W/M_Z.
\eeqn
The $\chi$ represents the Goldstone. We take the 't Hooft-Feynman
gauge with $\xi_W=\xi_Z=\xi_A=1$ so that no ``longitudinal" term
in the gauge propagators contributes. Not only does this make the
expressions much simpler and avoids unnecessary large
cancellations, but it also avoids the need for high tensor
structures in the loop integrals. The use of the five parameters,
$\tilde{\alpha}, \tilde{\beta}, \tilde{\delta}, \tilde{\kappa},
\tilde\varepsilon $ is not redundant as often these parameters
check complementary sets of diagrams. It is important  to note
that in order to successfully achieve this test one should not
include any width in the propagators. In fact our tree-level
results do not include any width. Because of the parameters and
the energies we consider, no width is required to regularise the
cross section. We verified that the inclusion of a width, at
tree-level, for the Higgs is minuscule. At one-loop order,
effective running widths are generated.

\noi {\it {\bf iv)}}Phase space integration for the $2 \to 2$
process is performed using the Gauss-Patterson method. While for
integration connected to $2 \to 2+\gamma$ processes, we use VEGAS
adaptive Monte Carlo integration package \cite{cuba}. For the
radiative process, we choose $k_c$ small enough  and verify  the
calculation for the total unpolarised cross section with the help
of {\tt CompHEP} \cite{CompHep}. We then check the stability and
independence of the result for the total and differential
corrections, Eq.~\ref{Eq:total_corr}, with respect to $k_c$.

%%%%%%%%%%%%%%%%%%%%%%%%%%%%%%%
\subsection{Extraction of some photonic corrections, genuine electroweak corrections}
%%%%%%%%%%%%%%%%%%%%%%%%%%%%%%%
The QED corrections contained in the electroweak corrections may
be important corrections however they do not reveal much about the
inner structure. We do not expect their contribution to be large
as in the case of the \epemt initiated processes which are
dominated by collinear singularities. This is because of the large
mass of the $W$ and top. Nonetheless one can attempt to extract
their effect, assuming such a split between weak and QED is
possible. For \wwttt this is not trivial because of  the
non-abelian nature of the photon radiation from the $W$. This said
the extraction of the leading log, $k_c$ dependent, contribution
in the virtual+soft can be considered as a {\em genuine} QED
correction. For \zzttt this QED correction can be extracted most
easily by a simple adaptation of the one loop initial state
radiation, ISR, in $e^+e^-$. We thus take, with $Q_t=2/3$ the
charge of the top, the relative QED correction to be:

\begin{equation}
\label{d_rv}
%\begin{split}
\delta^{qed,s+v}  =Q_t^2\frac {2\alpha}{\pi}  \left ( \frac
{1+\beta^2}{2\beta} \ln \frac {1+\beta}{1-\beta}-1 \right )\ln
\frac {2k_c}{\sqrt {s}}, \quad \beta = \sqrt{1- \frac{4m^2}{s}}.
\end{equation}

At threshold, $\beta \ra 0$, the QED corrections can be extremely
large. This is due to the Coulomb correction. This can be
extracted from:

\beqn \label{dcoul} \delta^{qed,s+v}_{{\rm Coul}}&=&Q_t^2 \frac
{2\alpha}{\pi} \frac {1+\beta^2}{2\beta}\left [ {\mbox {Sp}}\left
(\frac {2\beta}{\beta -1}\right )-{\mbox {Sp}}\left (\frac
{2\beta}{\beta + 1}\right )+\frac {\pi^2}{2}\right ], \quad Sp(z)
= -\int_0^z
dt\frac{Log(1-t)}{t} \nonumber \\
\delta^{qed,s+v}_{{\rm Coul}}( \beta \to 0 )&=& Q_t^2 \frac {\pi
\alpha}{2 \beta}.
\eeqn

This Coulombic correction is also present in \wwtt. The
corresponding QCD correction is obtained by $ Q_t^2 \alpha \ra C_F
\alpha_s$, with $C_F=4/3$ or in effect $\alpha \ra 3 \alpha_s$.

For \zzttt the ``purely" electroweak correction $(PEW)$ is defined
from the contribution of the soft and virtual corrections, {\it i.e.}
from $|\mathcal{M}_1|^{2,s+v}$ in Eq.~\ref{Eq:total_corr} and
therefore does not include hard photon radiation. To this we
subtract the QED correction given by Eq.~\ref{d_rv} so that:
\begin{equation}
\sigma_{ZZ}^{\scriptscriptstyle PEW} =
\sigma_{ZZ}^{\scriptscriptstyle s+v} - \delta^{qed,s+v}\;
\sigma_{ZZ}^{\scriptscriptstyle Born}=\delta^{PEW}
\sigma_{ZZ}^{\scriptscriptstyle Born}.
\end{equation}
This expression is gauge invariant, UV and IR finite and
independent of the photon energy cutoff.

For \wwtt, $\delta^{qed,s+v}$ involves radiation from the initial,
the final as well as their interference. Since we are only seeking
the $k_c$ dependent term, we decided to extract it numerically. To
achieve this we take 2 different values of the cut $k_c$, say
$k_{c1}$ and $k_{c2}$. Then, for the integrated cross section the
purely electroweak correction is defined as:

\beqn
\sigma_{WW}^{\scriptscriptstyle PEW} &=&
\sigma_{WW}^{\scriptscriptstyle s+v}(k_c) - A(\sqrt{s})\ln(2
k_c/\sqrt{s}), \quad A(\sqrt{s}) =
\frac{\sigma_{WW}^{\scriptscriptstyle
EW}(k_{c2})-\sigma_{WW}^{\scriptscriptstyle
EW}(k_{c1})}{\ln(k_{c2}/k_{c1})}.
\eeqn

\subsection{Genuine electroweak corrections in the $G_\mu$ scheme}
As stated earlier we work in a scheme where the electromagnetic
coupling, $\alpha$, is defined in the Thomson limit while our
processes occur at the TeV scale. The running $\alpha$ can induce
large corrections due  essentially to the light fermions. Moreover
some $m_t^2$ universal correction from $\Delta \rho$ can also be
induced. It is therefore more advantageous to work in the $G_\mu$
scheme. This helps extract the rather large universal corrections
contained in the two point function of the vector bosons
self-energies through $\Delta r$ which is of order $3\%$ for the
Higgs masses we are considering. In this scheme the electroweak
corrections are obtained by subtracting $2\,\Delta r$ from the
percentage change for the $PEW$ correction $\delta^{PEW}$ , hence
defining $\delta^{pew}=\delta^{PEW}-2 \Delta r$.

\

\section{One loop electroweak and QCD corrections:  results}
%%%%%%%%%%%%%%%%%%%%%%%%%%%%%%%

\subsection{Input parameters}
In getting our results we used the following parameters:
\begin{table}[H]
\begin{tabular}{l l l l l l l}
{\scriptsize $\alpha^{-1}=137.0359895$} &{\scriptsize$ M_Z=91.1875\, GeV$} &{\scriptsize$m_e=0.51099907\, MeV $}   &{\scriptsize$ m_\tau=1.777\, GeV $} & {\scriptsize$m_c=1.5\, GeV$} & {\scriptsize $m_t=173.7\, GeV$}\\
    &{\scriptsize $M_W=80.45\, GeV$}   &{\scriptsize $m_\mu=105.658389\, MeV$} &{\scriptsize$m_u=m_d=53.8\, MeV$}   &{\scriptsize $m_s=150\, MeV$} & {\scriptsize $m_b=4.7\, GeV$}
\end{tabular}
\end{table}

\noindent For $\alpha_s$ we use a running constant $\alpha_s(\mu)$
with $\mu=\sqrt{s_{VV}}$. The running is  evaluated at the
tree-loop level within the ${\overline{\rm MS}}$ with $5$
flavours and the normalisation $\alpha_s(M_Z)=0.1172$.
We have, for instance, $\alpha_s(\sqrt{s_{WW}}=500\,GeV) =
0.09432$ and $\alpha_s(\sqrt{s_{WW}}=1\,TeV) = 0.08776$. We also
set the CKM matrix to unity.

\subsection{Total cross section}
\begin{figure}[htbp]
\begin{center}
  \subfigure[$Z_LZ_L \to t\bar{t}$]{\epsfig{figure=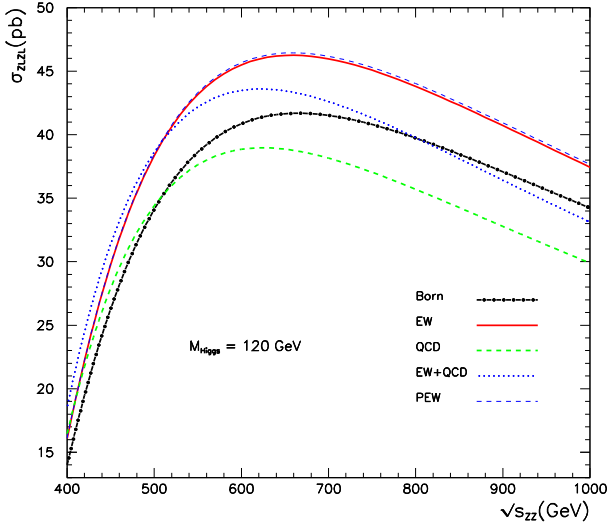,width=8.6cm}}
  \subfigure[$W_L^-W_L^+ \to t\bar{t}$ ]{\epsfig{figure=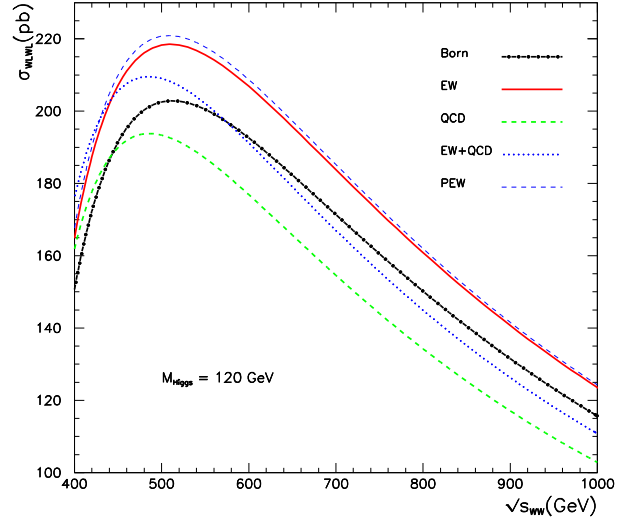,width=8.6cm}}
  \subfigure[$Z_LZ_L \to t\bar{t}$]{\epsfig{figure=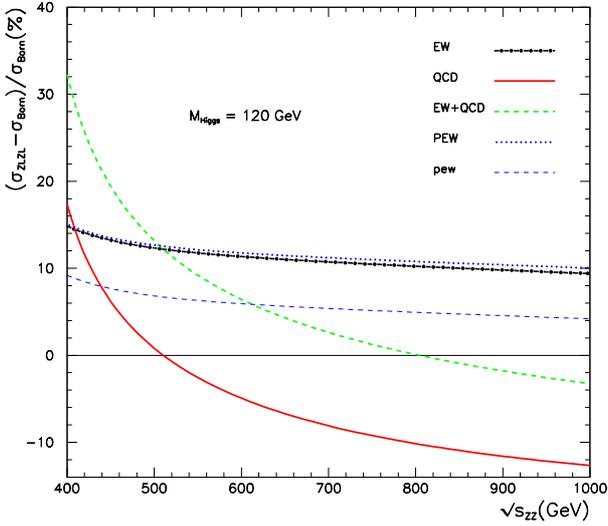,width=8.6cm}}
  \subfigure[$W_L^-W_L^+ \to t\bar{t}$ ]{\epsfig{figure=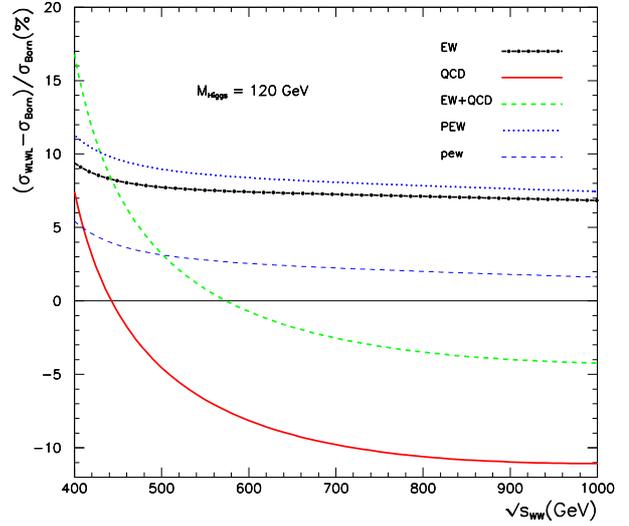,width=8.6cm}}
\end{center}
\caption{{\em Total cross section for the Born, the NLO with full
electroweak (EW), QCD,  purely electroweak (PEW) with the
extraction of the QED corrections and in the $G_\mu$ scheme(pew),
see text.  a) ${Z_LZ_L \to t\bar{t}}$ and b) ${W_L^-W_L^+ \to
t\bar{t}}$; ${M_{Higgs} =}$ 120 {\footnotesize GeV}. c) and d)
give the percentage change. }}\label{sig-zzwwtt-LLUU}
\end{figure}
The effect of the radiative corrections for ${M_{Higgs} = 120 \
GeV}$ is shown in figure \ref{sig-zzwwtt-LLUU} for the total cross
section for centre of mass energies of the $VV$ system ranging
from threshold to $1$TeV. Results for ${M_{Higgs} = 200,300 \,
GeV}$ are listed in Table~\ref{tableau}. First of all, let us
point to a common feature we found for all Higgs masses we
studied. The QED corrections we extracted are relatively quite
small so that there is little difference between the full EW
corrections including hard radiation and what we call the
``purely" EW corrections. This difference is largest, around $2\%$
at threshold energies (around $400$GeV) in the case of \wwttt.
Otherwise the difference is less than $1\%$ for both \zzttt and
\wwttp. Therefore in the remainder of our discussion we will refer
to the EW corrections having in mind the full EW. For both $ZZ$
and $WW$, we find large positive corrections at the lower
energies. As we discussed earlier, this is due, to a large extent,
to the Coulomb correction. Apart from the low energy region, the
corrections we find for the case of the $WW$ and $ZZ$ are within $3\%$ of each other, be it for the EW or QCD
corrections. At the threshold region though, the QCD corrections
are large, especially for $ZZ$ and increase somehow as the Higgs
mass is increased from $120$GeV to $300$GeV. Past the threshold
region the QCD corrections decrease quite rapidly turning negative
with a correction of about $-10\%$ at $\sqrt{s_{VV}}=1$TeV. In
contrast, past the threshold region, the electroweak corrections
decrease extremely slowly and are almost constant with a value of
around $10\%$. Especially at high energies, the electroweak
corrections tend therefore to cancel out most of the QCD
corrections. The electroweak corrections should therefore be taken
into account on par with the QCD corrections. Of course,
expressing the tree-level cross section in terms of $G_\mu$ does
account for a large part of the EW corrections. In this scheme the
corrections are in the range $2-5\%$, slightly larger for $ZZ$
with ${M_{Higgs} = 120 \, GeV}$. However even in this scheme the
corrections can be competitive with the QCD corrections as the
latter become small in the energy range $\sqrt{s}=500-700$ GeV.

\begin{table}[H]
\begin{center}
\begin{tabular}{|c|c|c|c|c|c||c|c|c|c|c|} \hline
$W_L^-W_L^+ \to t\bar{t}$ &\multicolumn{5}{c||}{${M_{Higgs} =}$ 200\,{\footnotesize GeV}} &\multicolumn{5}{c|}{${M_{Higgs} =}$ 300\,{\footnotesize GeV}} \\\hline
${\scriptstyle \sqrt{s_{WW}}\,(GeV)}$ & $\sigma_0$ {\scriptsize (pb)} & $\delta^{EW}$ &$\delta^{PEW}$ & $\delta^{pew}$  & $\delta^{\scriptsize QCD}$ & $\sigma_0$ {\scriptsize (pb)} & $\delta^{EW}$ &$\delta^{PEW}$ & $\delta^{pew}$  & $\delta^{\scriptsize QCD}$ \\ \hline\hline
400  & 144.33  & 9.35  & 11.17 & 4.97 &  6.44 & 163.77  & 10.25 & 11.93 & 5.42 & 10.54\\\hline
500  & 196.13  & 8.23  &  9.43 & 3.23 & -4.99 & 201.23  &  9.41 & 10.52 & 4.00 & -3.97\\\hline
600  & 188.49  & 7.96  &  8.91 & 2.72 & -8.33 & 189.66  &  8.93 &  9.82 & 3.30 & -7.89\\ \hline
700  & 168.84  & 7.79  &  8.61 & 2.41 & -9.87 & 169.04  &  8.67 &  9.43 & 2.92 & -9.59\\\hline
800  & 148.57  & 7.63  &  8.36 & 2.16 & -10.61& 148.52  &  8.43 &  9.12 & 2.60 &-10.42 \\\hline
900  & 130.46  & 7.48  &  8.15 & 1.95 & -10.96& 130.36  &  8.22 &  8.84 & 2.33 & -10.81 \\ \hline
1000 & 114.95  & 7.33  &  7.95 & 1.75 & -11.07 & 114.84  &  8.01 &  8.59 & 2.08 & -10.93  \\ \hline \hline\hline

$Z_LZ_L \to t\bar{t}$ &\multicolumn{5}{c||}{${M_{Higgs} =}$ 200\,{\footnotesize GeV}} &\multicolumn{5}{c|}{${M_{Higgs} =}$ 300\,{\footnotesize GeV}} \\\hline
${\scriptstyle \sqrt{s_{ZZ}}\,(GeV)}$ & $\sigma_0$ {\scriptsize (pb)} & $\delta^{EW}$ &$\delta^{PEW}$ & $\delta^{pew}$  & $\delta^{\scriptsize QCD}$ & $\sigma_0$ {\scriptsize (pb)} & $\delta^{EW}$ &$\delta^{PEW}$ & $\delta^{pew}$  & $\delta^{\scriptsize QCD}$ \\ \hline\hline
400  & 21.87  &  9.96 & 10.14 & 3.94 & 22.46 & 98.10  & 10.18 & 10.36 & 3.85 & 30.87 \\\hline
500  & 44.24  & 10.13 & 10.46 & 4.26 &  2.93 & 96.56  & 10.19 & 10.51 & 3.99 &  7.9\\\hline
600  & 48.49  & 10.15 & 10.57 & 4.37 & -4.06 & 78.83  & 10.21 & 10.62 & 4.10 & -1.42\\ \hline
700  & 46.89  & 10.04 & 10.53 & 4.33 & -7.83 & 65.39  & 10.72 & 11.20 & 4.68 & -6.55\\\hline
800  & 43.53  &  9.84 & 10.38 & 4.18 & -10.17& 55.45  & 10.78 & 11.32 & 4.80 & -9.69  \\\hline
900  & 39.81  &  9.57 & 10.16 & 3.96 & -11.77& 47.85  & 10.61 & 11.20 & 4.68 & -11.74 \\ \hline
1000 & 36.23  &  9.27 & 9.91  & 3.71 & -12.91& 41.87  & 10.33 & 10.96 & 4.44 & -13.14  \\  \hline

\multicolumn{6}{c}{}
\end{tabular}
\caption{{\em The Born total cross section $\sigma_0$ and the
relative corrections for the electroweak ({$\delta^{\footnotesize
EW}$}), ``purely" electroweak ({$\delta^{\footnotesize PEW}$}),
the $G_\mu$ scheme ({$\delta^{\footnotesize pew}$}) and QCD
({$\delta^{\footnotesize QCD}$}). \label{tableau}}}
\end{center}
\end{table}

%%%%%%%%%%
\subsection{One loop angular distributions}
\begin{figure}[H]
\begin{center}
\subfigure[$K_{WW}^{\scriptscriptstyle SM}$ rates for
$\sqrt{s_{WW}}= 400\, GeV$
  \label{kanomrate400}]{\epsfig{figure=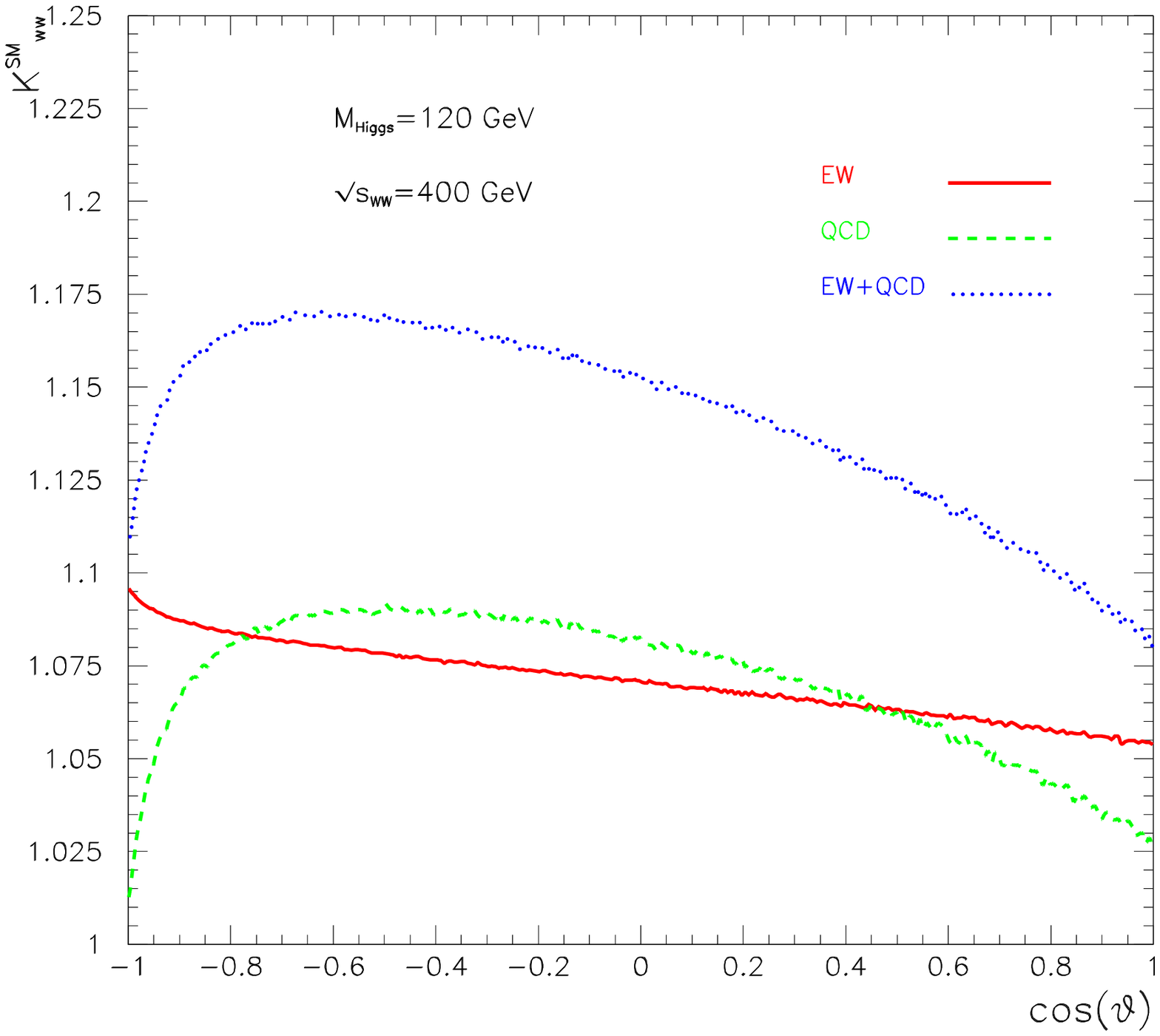,width=8.6cm}}
  \subfigure[$K_{WW}^{\scriptscriptstyle SM}$ rates for $\sqrt{s_{WW}}= 1000\, GeV$
  \label{kanomrate1000}]{\epsfig{figure=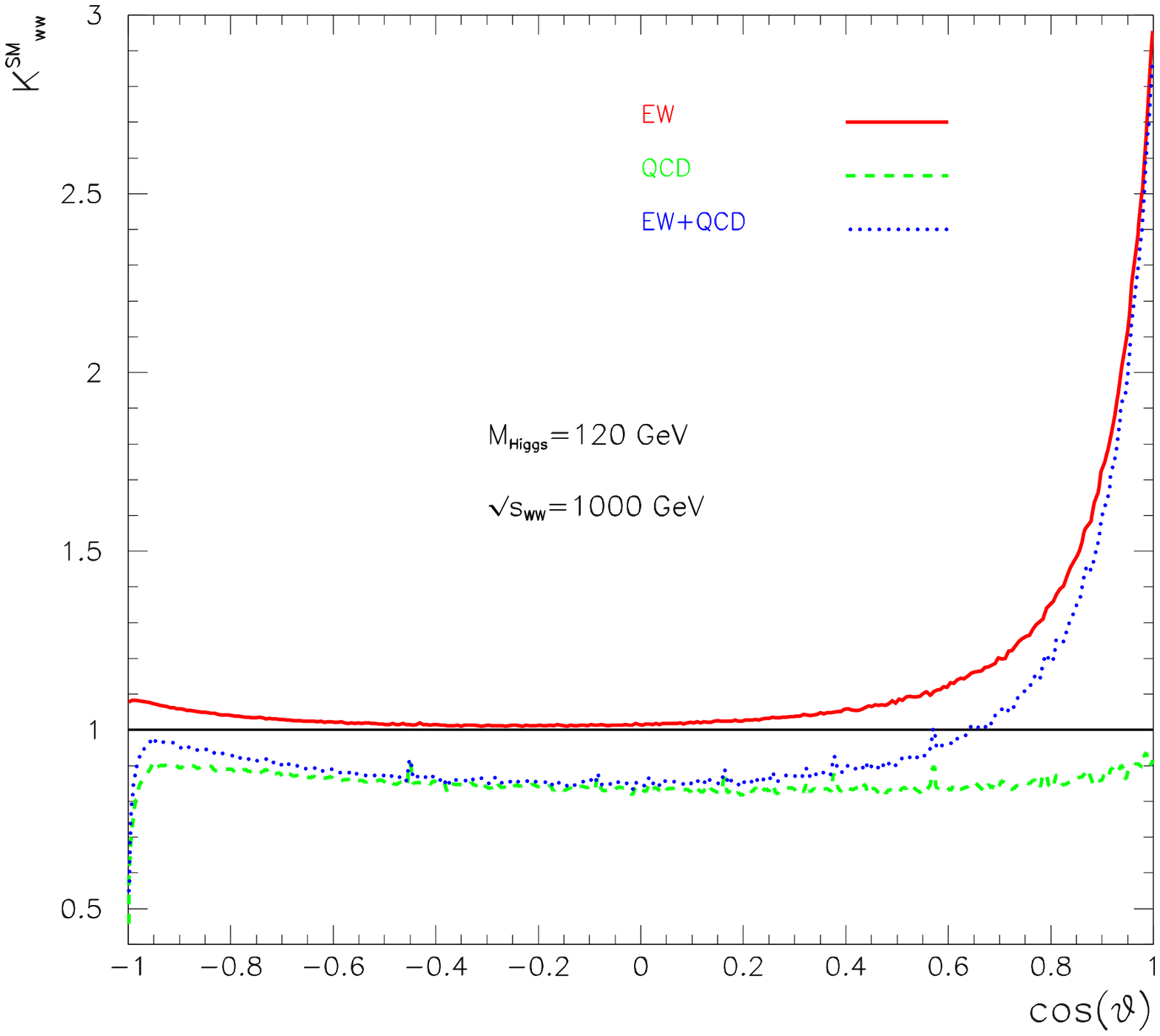,width=8.6cm}}
\end{center}
\caption{{\em The correction $K_{WW}^{\scriptscriptstyle SM}$
relative to the \sm for the unpolarised cross section $W_U^-W_U^+
\to t\bar{t}$ with $M_{Higgs} = 120\, GeV$, (a) $\sqrt{s_{WW}}=
400\, GeV$,(b) $\sqrt{s_{WW}}= 1000\, GeV$. The \sm Born
distributions are dispalyed in Fig.~\ref{diffcross-tree}. }}
\label{sigkww-zzwwtt-LLUU}
\end{figure}

We show the corrected angular distributions for an energy not far
from threshold, $\sqrt{s_{VV}} = 400$ GeV, and the other at the high energy of
$1$TeV. The corresponding $K_{WW}^{\scriptscriptstyle SM}$ factors
for the unpolarised case, defined to be the ratio of the one loop corrected
differential cross section to the corresponding tree level one,
are shown in  Fig.~\ref{sigkww-zzwwtt-LLUU}. In the lower energy where the 
cross section is largest, see Fig.~\ref{diffcross-tree},
the electroweak correction dominates over the QCD  one.
 In the central region the two
corrections are both positive and of very similar size. At high
energy, the electroweak correction shows a very large effect in
the forward region, while it is quite modest in the backward
region and almost zero in the central region. The large correction
in the forward region has to be put in perspective though. Indeed,
in this region the differential cross section is 4 orders of
magnitude smaller than the in the backward region. This is just an
effect due to hard photon radiation that migrates some of the
events from the most populated region. In the central region the
electroweak correction is almost nil. The QCD correction on the
other hand is negative all throughout but even more so in the
backward direction. This is at the origin of the relatively large
and  negative correction at the level of the integrated cross
section.
%%%%%%%%%

%%%%%%%%%%
\section{Results at $e^+ e^-$ level}
%%%%%%%%%%%%

\begin{figure}[hbtp]
\begin{center}
  \subfigure[$K_{ZZ}$ factor for $e^+e^- \to t\bar{t}e^+e^-$ through $
  {Z_LZ_L \to t\bar{t}}$ \label{k-ilc-zz}]{\epsfig{figure=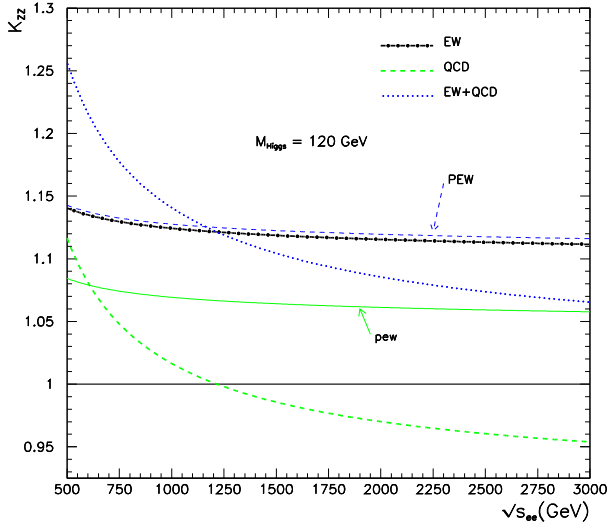,width=8.6cm}}
  \subfigure[$K_{WW}$ factor for $e^+e^- \to t\bar{t}\nu_e\bar{\nu_e}$ through ${W_LW_L \to t\bar{t}}$
  \label{k-ilc-ww}]{\epsfig{figure=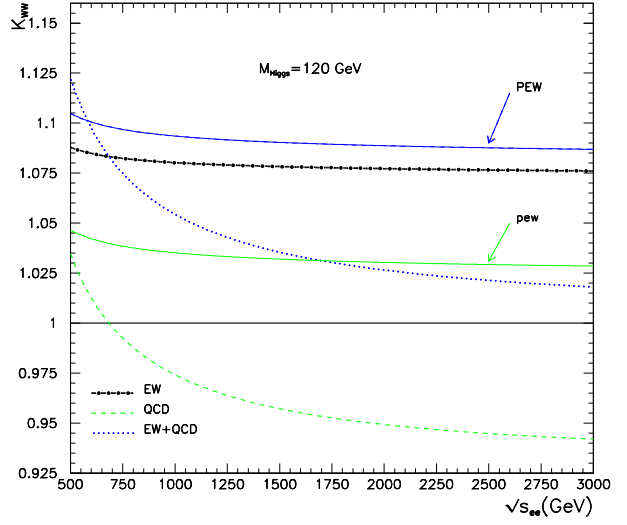,width=8.6cm}}
 \subfigure[$K_{ZZ}$ factor for $e^+e^- \to t\bar{t}e^+e^-$ through ${Z_LZ_L \to t\bar{t}}$
 \label{k-ilc-zz-MH300}]{\epsfig{figure=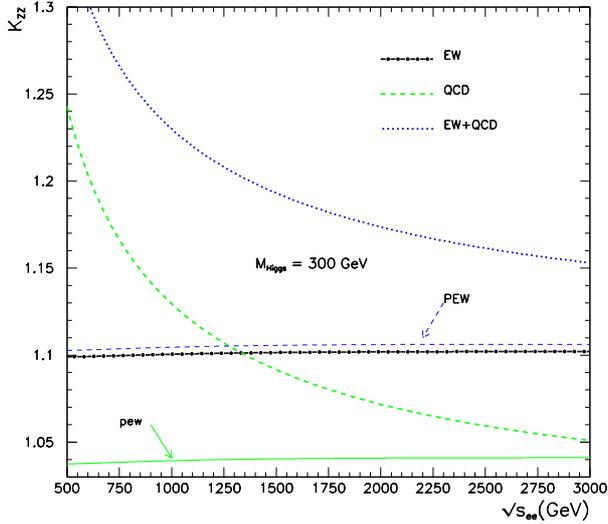,width=8.6cm}}
  \subfigure[$K_{WW}$ factor for $e^+e^- \to t\bar{t}\nu_e\bar{\nu_e}$ through
  ${W_LW_L \to t\bar{t}}$ \label{k-ilc-ww-MH300}]{\epsfig{figure=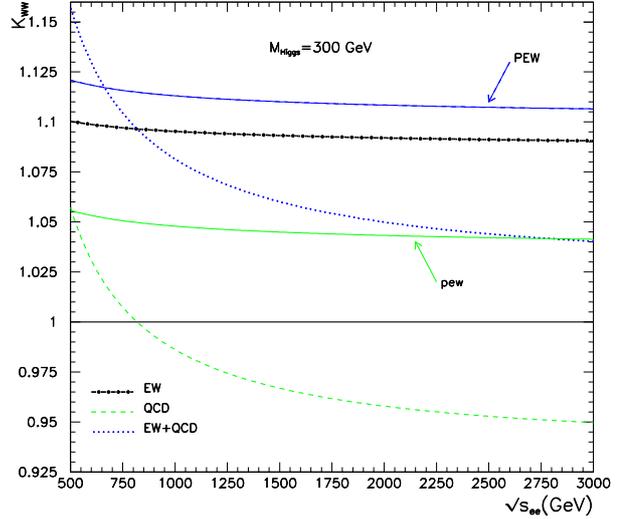,width=8.6cm}}
\end{center}
\caption{{\em $K$ factor representing the loop corrections to
(a,c) $e^+e^- \to t\bar{t}e^+e^-$ through ${Z_LZ_L \to t\bar{t}}$
and (b,d)$ e^+e^- \to t\bar{t}\nu\bar{\nu}$ process through
${W_LW_L \to t\bar{t}}$. (a,b) are for $M_{Higgs}=120$GeV and
(c,d) for $M_{Higgs}=300$GeV.}} \label{k-ilc-zz-ww}
\end{figure}

We now convolute the results found at the ``partonic" level $VV
\ra t \bar t $ to $\epem \ra t \bar t + X$. As discussed in the
previous sections we impose  the kinematical cut $m_{t\bar{t}}>
400\, GeV$ together with $p_T^{t,\bar t}>10$GeV. We have
considered a centre-of mass energy for the linear collider ranging
from $500$GeV to $3$TeV. As we discussed previously, for
$\sqrt{s_{ee}}<1$TeV, the EVBA approximation for the process should
be taken with a grain of salt. We do this however in order  to
also comply with the analysis conducted in \cite{GZ05} and be able
to compare the results for the QCD corrections we find with
theirs. As in \cite{GZ05} we quantify the relative correction at
the \epemt level as:
\begin{equation}
K_{VV}=(\sigma_{VV}^{NLO})/\sigma_{VV}^{LO}\equiv 1+\delta^{NLO}.
\end{equation}

Figure~\ref{k-ilc-zz-ww} shows the corrections  for $M_{Higgs}=120
GeV$ and $M_{Higgs}= 300 GeV$ as a function of the centre-of-mass
energy, $\sqrt{s_{ee}}$. Our results for the $EW$, $PEW$ and $pew$
corrections have a straightforward interpretation if one recalls
the results obtained at the $VV$ level. Indeed, we found that the
EW corrections, as the $VV$ energy is varied, is almost constant,
with roughly the same correction for the $WW$ and $ZZ$ channel for
the Higgs masses we studied. The correction amounts to about
$10\%$, or to about $5\%$ in the $G\mu$ scheme. No wonder that
this is roughly what we get after convolution. For the QCD part,
the corrections depend on $\sqrt{s_{ee}}$. For low
energies,$\sqrt{s_{ee}}=500$GeV, the QCD corrections for $ZZ$ are
of order $10\%$ for $M_{Higgs}=120$GeV and almost twice as much
for $M_{Higgs}=300$GeV. They do however drop quickly being
sensibly equal to those found in $WW$, turning negative of order
$-5\%$ at $\sqrt{s_{ee}}=3$TeV. These results can again be guessed
based on what we found at the $VV$ level. We completely agree with
the results of\cite{GZ05} for the QCD corrections, at least for
$M_{Higgs}=120 GeV$ that we both consider. To summarise, we see
that for $\sqrt{s_{ee}}=1.5$TeV were these processes are most likely
to be useful and where the EVBA can be trusted, in the $WW$
channel the electroweak corrections are of order $10\%$ while QCD
correction is of order $-5\%$. Even translating the electroweak
results in the $G\mu$ scheme, the electroweak corrections are of
order $4\%$ and practically wash out the QCD corrections.

\section{Effects of  anomalous top quark interactions}
%%%%%%%%%
%%%%----
As we argued in the introduction, the processes we are studying
might be a powerful tool to reveal novel interactions of the top
quarks to the vector bosons, in particular their longitudinal
modes. Without picking up any specific dynamical model in
particular, a simple way of studying such effects is to revert to
the effective Lagrangian approach. The most general lowest order
operators, beyond those in the \sm have been given
in\cite{LariosYuan}. We will only pick up two such operators and
see their effect in \wwtt. In the unitary gauge and picking up
only the $WWt \bar t$ couplings one can write:
\begin{equation}
\label{anomalop}
\mathcal{L}_{eff}= \mathcal{L}_{eff}^{(a_1)} +
\mathcal{L}_{eff}^{(a_2)} = \frac{a_1}{\Lambda} \bar{t} g^{\mu
\nu}t {W^+}_{\mu}{W^-}_{\nu} + \frac{a_2}{\Lambda} \bar{t}
(i{\sigma}^{\mu \nu})t {W^+}_{\mu}{W^-}_{\nu},
\end{equation}
$a_1$ is a scalar interaction which contribute to the $S$-wave and
$a_2$ is a magnetic interaction which contributes to the $P$-wave,
see Appendix for the helicity amplitudes. $\Lambda$ is the scale
of New Physics associated with symmetry breaking which we take in
accordance with\cite{LariosYuan} to be $\Lambda=4\pi v\sim
3.1$TeV. Reference\cite{LariosTait} finds that $a_{1,2}$ of order
$0.1$ can be probed at a linear collider with
$\sqrt{s_{ee}}$=1.5TeV and an integrated luminosity of
$200$fb$^{-1}$ in precisely the $W$ fusion process we are
studying. These bounds may be improved with the use of top
polarisation\cite{LariosYuan}.

\subsection{Effects in \wwtt}
\begin{figure}[htbp]
\begin{center}
  \subfigure[Differential distributions for  $\sqrt{s_{WW}}= 400\, GeV$
  \label{anomewqcd400}]{\epsfig{figure=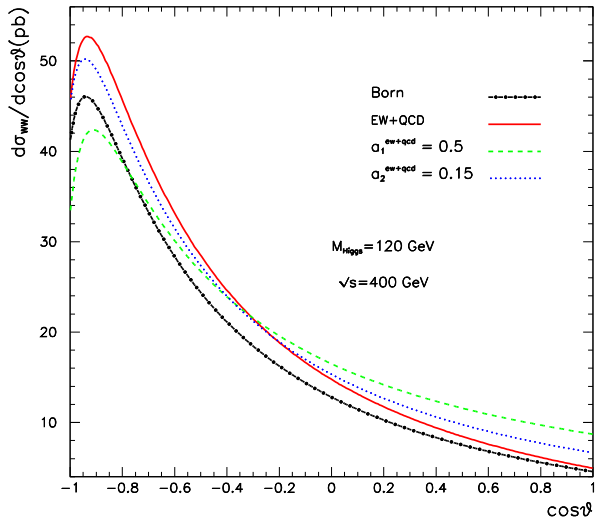,width=8.6cm}}
  \subfigure[Differential distributions for  $\sqrt{s_{WW}}= 1000\, GeV$
  \label{anomewqcd1000}]{\epsfig{figure=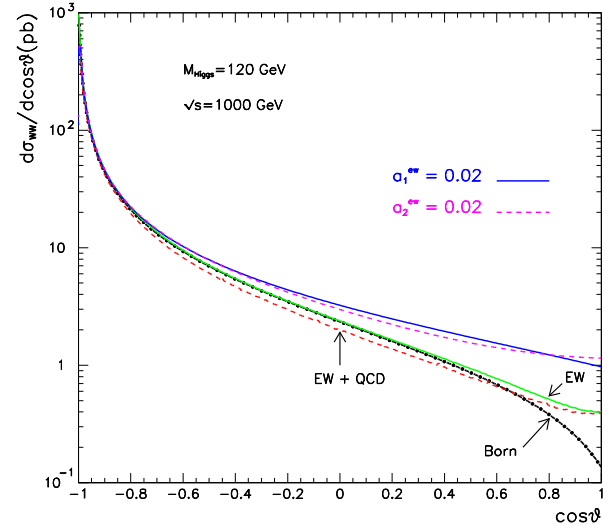,width=8.6cm}}
\end{center}
\caption {{\em Angular distributions for the SM at tree and loop
level and those including the anomalous couplings that give the
same total cross section as the corrected total cross section, see
text. The figures are for the unpolarised cross section.
$M_{Higgs} = 120\, GeV$, for  (a) $\sqrt{s_{WW}}= 400\, GeV$, (b)
$\sqrt{s_{WW}}= 1000\, GeV$} }
\label{anomewqcdfull-wwtt-UUUU}
\end{figure}

The $a_i$ we consider will be of the order $0.1$ or lower as
indicated by the bounds set at an $1.5$TeV
collider\cite{LariosTait}. They will contribute essentially through
an interference effect with the \sm (tree-level) amplitude
contribution. Apart from their effects on the total cross section,
we wish to see how they affect the distributions. This not only
can tell apart the contribution of the two operators but also
differentiate between the new operators and the \sm prediction.
Taking into account the \sm loop corrections one can ask whether
the operators can mimick the loop corrections. For this purpose we
set the value of the $a_i$ such that it reproduces the one-loop
{\em corrected} \sm value for the unpolarised total cross section. For
example, $a_1^{ew}$ is the value of $a_1$ that reproduces the full
electroweak correction,
\begin{equation}
\sigma^{anom1ew}(a_1^{ew}) = \sigma_{1loop}^{SM}(EW).
\end{equation}
Likewise, $a_{i}^{qcd}$ will represent the QCD correction and
$a_{i}^{ew+qcd}$ the total one-loop correction. We restrict
ourselves to $M_{{\rm Higgs}}=120$GeV. For example, for
$\sqrt{s_{WW}}= 400\, GeV$, we find $a_1^{ew+qcd} = 0.5$ and
$a_2^{ew+qcd} = 0.15$ while for $\sqrt{s_{WW}}= 1000\, GeV$, we
get $a_{1,2}^{ew} = 0.02$ an order of magnitude smaller, but
perhaps too optimistic from the experimental measurement, at least
for $1.5$TeV collider, but not necessarily for a $3$TeV machine.
When we look at the angular distributions, see
Fig.~\ref{anomewqcdfull-wwtt-UUUU}, we see as expected, that in
principle one can tell the different effects apart, especially for
low energies. At 1TeV $WW$ centre of mass, one needs to get to the
central region, where unfortunately the cross section is much
smaller. Nonetheless, we clearly see that one should take into
account the effect of loop corrections when looking for new
effects.

\subsection{Effects in \wwtt \,at \epemt level: convolution}
We now convolute with the $e/W$ distribution functions for the
longitudinal mode and weigh the effect of the anomalous couplings
with those of the radiative corrections as a function of the
\epemt centre of mass energy for $e^+e^- \to
t\bar{t}\nu_e\bar{\nu_e}$. For illustration we take
$a_1=a_2=0.03$. We look at the effect on, the ``$K$-factor"
$K_{WW}=\sigma^{loop,a_i}/\sigma^{{\rm Born}}$.

As expected, the effect due to the anomalous couplings increases
with energy. The lesson one also gets is that though one can
measure a discrepancy with a tree-level calculation due to the
anomalous interaction, this is not necessarily the case compared
to the use of a corrected \sm cross section. Moreover, ``in real
life" an anomalous contribution might be there but its effect is
washed out or dramatically reduced due to the ``contamination"
from the QCD or the electroweak corrections, if these are not
taken into account. This said, in ``real life", other important
issues must be taken care of such as the background and the use of
a calculation that can go beyond the EVBA approximation. However,
the study we have made here clearly shows that taking into account
both the electroweak and QCD corrections is an important issue.
%%%%%%%%%%%%
\begin{figure}[H]
\begin{center}

\mbox{\includegraphics[width=0.45\textwidth,height=0.45\textwidth]
{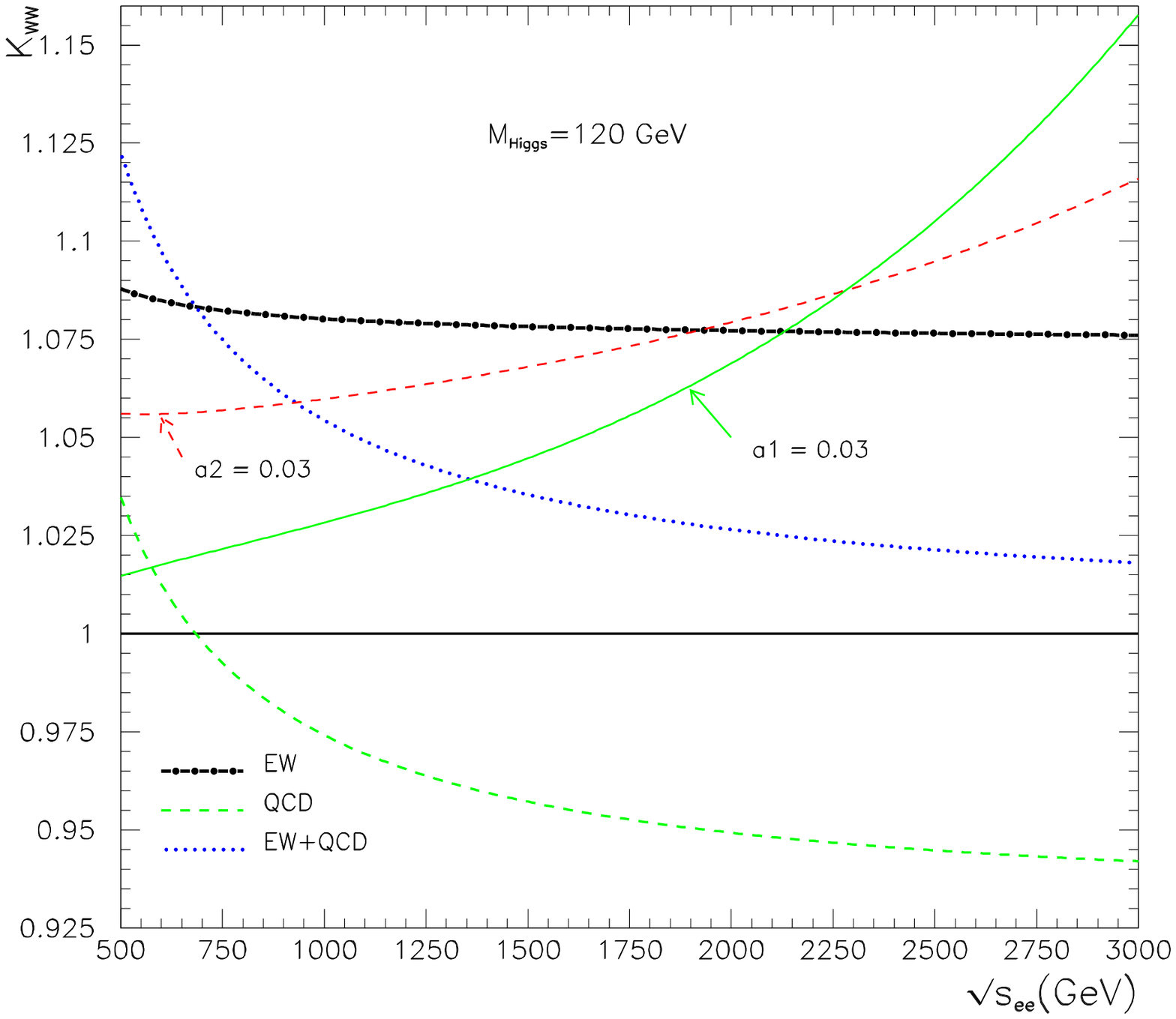}\includegraphics[width=0.45\textwidth,height=0.45\textwidth]{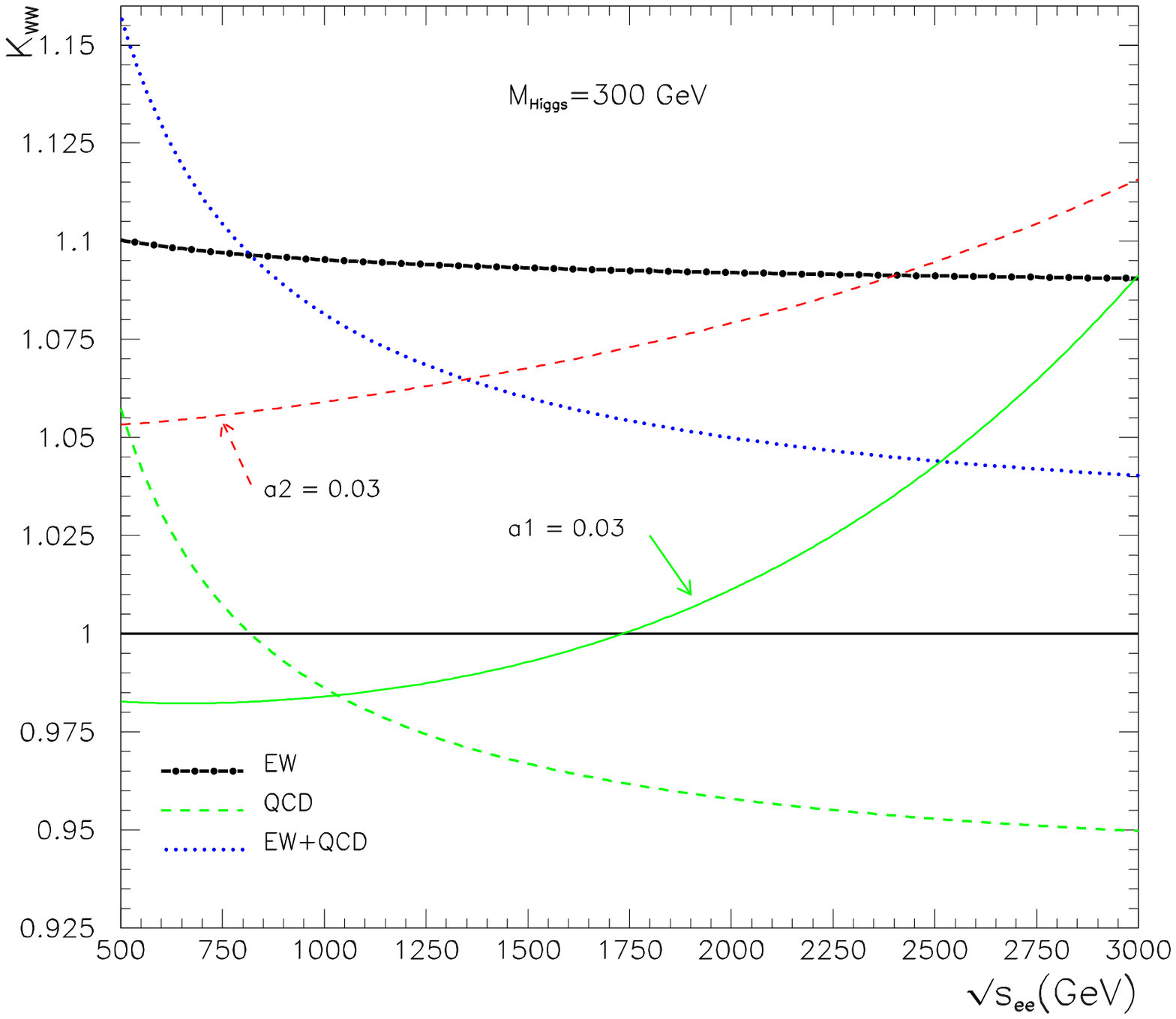}}
\caption{{\em $K_{WW}$  with $a_1=a_2=0.03$ for $e^+e^- \to
t\bar{t}\nu_e\bar{\nu_e}$ through ${W_LW_L \to t\bar{t}}$ compared
to the one-loop $K$-factors.}} \label{k-ilc-ww}

\end{center}
\end{figure}

%%%%%%%%%%%%%%%%%%%%%%%%%%%%%%%%%%%%%%%%%%%%%%%%%%%%%%%%%%
\section{Conclusion}
%%%%%%%%%%%%%%%%%%%%%%%%%%%%%%%%%%%%%%%%%%%%%%%%%%%%%%%%%%%
There are plenty of arguments that point to the top playing a very
special role in the mechanism of symmetry breaking. \wwttt and
\zzttt at high energy provide a very nice testing ground for
revealing features of symmetry breaking that may not be probed
otherwise since they bring together the top and the Goldstone
bosons. On the other hand within the \sm and because of the large
Yukawa coupling of the top, the electroweak virtual effects of the
top are often large. We have performed a full electroweak
correction to these processes together with the QCD corrections.
The latter had been calculated recently and we agree with their
findings. We  find that the electroweak corrections can be large
of order $10\%$, though smaller of order $5\%$ if we use the
$G_\mu$ scheme. However, even in this scheme the electroweak
corrections can be of order the QCD corrections and combine such
that the two corrections cancel each other for a certain range of
energy. The combined effect should therefore be taken into
consideration when looking for New Physics effects in these
processes. To illustrate the role of the radiative corrections,
both QCD and electroweak, we have also considered the effect of
some anomalous $W$-top interactions. Unless these are substantial
enough to give deviations larger than $20\%$ or so, one needs to
take into account the radiative corrections. The latter can hide
the effects or even increase it. Naturally these important
processes can only occur as subprocesses at the LHC or the ILC.
For the LHC, QCD top production and the QCD backgrounds are such
as to prevent a study of these reactions. At the ILC, we have,
like others, relied on the EVBA to study these reactions and
therefore convoluted with  the electron structure function
describing the $W$ content of the electron. Still in this context
the interplay between the electroweak and QCD corrections is
important, even though for the Higgs masses we considered the
convolution for \epemt centre of mass energy in the TeV range
leads to the QCD and electroweak cancellling each other. Therefore
taking into account the QCD corrections might not be enough to
draw a meaningful conclusion in case a deviation is observed. Of
course, to draw more meaningful conclusions we would need, beside
a careful simulation with backgrounds, to attack the full
radiative correction to $\epem \ra t \bar t \nu \bar{\nu}$, which
is a formidable task. This said the gist of the electroweak
corrections should be contained at the level of the \wwtt. In this
study we restricted ourselves to a light enough Higgs as allowed
by the precision data. It would be interesting to review the
situation in the case of a very heavy Higgs or a manifestation of
the \sm without a Higgs.
\begin{center}
%%%%%%%%%%%%%%%%%%%%%%%%%%%%%%%%%%%%%%%%%%%%%%%%%%%%%%%%%%%
\section*{AKNOWLEDGMENTS}
%%%%%%%%%%%%%%%%%%%%%%%%%%%%%%%%%%%%%%%%%%%%%%%%%%%%%%%%%%%
\end{center}
One of us (N.B) is extremely grateful to LAPTH for its kind
hospitality during the whole period this work has been conducted
and to the Algerian Ministry of High Education and Sientific
Research for providing a grant (BAF/20051179). N.B also wishes to
thank Vincent LAFAGE for helpful discussions and for providing
codes on helicity amplitude methods with massive particles.

%%%%%%%%%%%%%%%%%%%%%%%%%%%%%%%%%%%%%%%%%%%%%%%%%%%%%%%%%%%
%%%%%%%%%%%%%%%%%%%%%%%%%%%%%%%%%%%%%%%%%%%%%%%%%%%%%%%%%%%
\renewcommand{\thesection}{\Alph{section}}
\setcounter{section}{0}

\section*{Appendix: Tree-level helicity amplitudes for \wwtt}

\renewcommand{\theequation}{\thesection A \arabic{equation}}
\setcounter{equation}{0}

\noi Let $\lambda_1=\pm{1}$ represent the transverse modes for the
incoming $W$ bosons, $0$ the corresponding longitudinal mode and
$\lambda_2=\pm{1}$ the top quark polarisations. For the top quark
the third component of the isospin component is $I_3 =
\frac{1}{2}$ and its charge $Q=\frac{2}{3}$. We also define $\gamma_i^2 = 1/(1- \beta_i^2) = s/4 M_i^2$ and $\pi_i$ the
reduced propagator ($\pi_i = s\,\Pi_i$), where $ \sqrt{s}$ stand
for the center of mass energy of the $WW$ system and
\begin{eqnarray}
\Pi_{\gamma}=\frac{1}{s} , \quad
\Pi_{Z}=\frac{1}{{s-M_{Z}^2}},\quad
\Pi_{H}=\frac{1}{s-M_{Higgs}^2},\quad \Pi_b =
\frac{-4}{s(\beta_t^2  + \beta_W^2)+4 m_b^2 + 2 s \beta_t \beta_W
cos\theta}
\end{eqnarray}
$\theta$ is the angle between the $W^-$ and the $t$ quark.

\noi The tree level helicity amplitudes for the \sm are given by
$\mathcal{M}^{\scriptscriptstyle
SM}_{\lambda_1,\lambda_2,\lambda_3,\lambda_4}$ for the process
$W^-(\lambda_1) W^+(\lambda_2) \to t(\lambda_3)
\bar{t}(\lambda_4)$
\begin{eqnarray}
\mathcal{M}_{\lambda_1,-\lambda_1,-\lambda_2,\lambda_2}^{\scriptscriptstyle
SM}=
-\lambda_1\lambda_2\frac{g^2}{8}\pi_b\beta_t(1-\lambda_2\beta_t)
(1-\lambda_1\lambda_2\cos\theta)\sin\theta
\end{eqnarray}

\begin{equation}
\mathcal{M}_{\lambda_1,-\lambda_1,\lambda_2,\lambda_2}^{\scriptscriptstyle
SM}=\lambda_1\frac{g^2}{8}\pi_b\frac{\beta_t}{\gamma_t}\sin^2\theta
\end{equation}

\begin{eqnarray}
\mathcal{M}_{\lambda_1,\lambda_1,-\lambda_2,\lambda_2}^{\scriptscriptstyle
SM}=\frac{g^2}{4}\sin\theta \Big
[\frac{\pi_b}{2}(1-\lambda_2\beta_t)(\beta_W+\beta_t\cos\theta) +
2\pi_Z(I_3(1-\lambda_2\beta_t)-2Q\frac{M_Z^2}{s}\sin^2\theta_W)\beta_W \Big
]
\end{eqnarray}

\begin{eqnarray}
\mathcal{M}_{\lambda_1,\lambda_1,\lambda_2,\lambda_2}^{\scriptscriptstyle
SM}=-\frac{g^2}{4}\frac{1}{\gamma_t}\Big
[\lambda_1\frac{\pi_b}{2}(\beta_W+\beta_t\cos\theta)
(1+\lambda_1\lambda_2\cos\theta) + \lambda_2\pi_H\beta_t \\
+ 2\lambda_2\pi_Z(I_3-2Q\frac{M_Z^2}{s}\sin^2\theta_W)\beta_W\cos\theta \Big
]
\end{eqnarray}

\begin{eqnarray}
\mathcal{M}_{0,\lambda_1,-\lambda_2,\lambda_2}^{\scriptscriptstyle
SM}=
\lambda_1\frac{g^2}{2\sqrt{2}}\gamma_W(1+\lambda_1\lambda_2\cos\theta)
\Big
[\frac{\pi_b}{4}(1-\lambda_2\beta_t)(\beta_W\beta_t+\lambda_1\beta_t-\lambda_2\beta_W
+\lambda_1\lambda_2\beta_W^2 \\ -2\lambda_2\beta_t\cos\theta) -
2\lambda_2\pi_Z(I_3(1-\lambda_2\beta_t)-2Q\frac{M_Z^2}{s}\sin^2\theta_W)\beta_W
\Big ]
\end{eqnarray}

\begin{eqnarray}
\mathcal{M}_{-\lambda_1,0,-\lambda_2,\lambda_2}^{\scriptscriptstyle
SM}=\mathcal{M}_{0,\lambda_1,-\lambda_2,\lambda_2}^{\scriptscriptstyle
SM}
\end{eqnarray}

\begin{eqnarray}
\mathcal{M}_{0,\lambda_1,\lambda_2,\lambda_2}^{\scriptscriptstyle
SM}= -\sqrt{2}\frac{g^2}{4}\frac{\gamma_W}{\gamma_t}\sin\theta
\Big [\frac{\pi_b}{4}(\beta_W\beta_t+\lambda_1\beta_t+\lambda_2\beta_W - \lambda_1\lambda_2\beta_W^2+2\lambda_2\beta_t\cos\theta)\\
+2\lambda_2\pi_Z(I_3-2Q\frac{M_Z^2}{s}\sin^2\theta_W)\beta_W \Big ]
\end{eqnarray}

\begin{eqnarray}
\mathcal{M}_{-\lambda_1,0,\lambda_2,\lambda_2}^{\scriptscriptstyle
SM}=\mathcal{M}_{0,\lambda_1,\lambda_2,\lambda_2}^{\scriptscriptstyle
SM}
\end{eqnarray}

\begin{eqnarray}
\mathcal{M}_{0,0,-\lambda_2,\lambda_2}^{\scriptscriptstyle
SM}=\frac{g^2}{4}\gamma_W^2\sin\theta
\Big [\frac{\pi_b}{2}(1-\lambda_2\beta_t)(2\lambda_2\beta_W\beta_t-\beta_W(1-\beta_W^2)-2\beta_t\cos\theta)\\
-
2\pi_Z(I_3(1-\lambda_2\beta_t)-2Q\frac{M_Z^2}{s}\sin^2\theta_W)(3-\beta_W^2)\beta_W
\Big ]
\end{eqnarray}

\begin{eqnarray}
\mathcal{M}_{0,0,\lambda_2,\lambda_2}^{\scriptscriptstyle
SM}=-\lambda_2\frac{g^2}{4}\frac{\gamma_W^2}{\gamma_t} \Big
[\frac{\pi_b}{2}((1+\beta_W^2)\beta_t-(1-\beta_W^2)\beta_W\cos\theta
- 2\beta_t\cos^2\theta) \nonumber \\ + \pi_H(1+\beta_W^2)\beta_t
-2\pi_Z(I_3-2Q\frac{M_Z^2}{s}\sin^2\theta_W)(3-\beta_W^2)\beta_W\cos\theta
\Big ]
\end{eqnarray}

The anomalous helicity amplitudes including the operators in
Eq.~\ref{anomalop} we denote by $\mathcal{N}$

\begin{equation}
\mathcal{N}_{\lambda_1,-\lambda_1,-\lambda_2,\lambda_2}=
\mathcal{N}_{\lambda_1,-\lambda_1,\lambda_2,\lambda_2}=
\mathcal{N}_{\lambda_1,\lambda_1,-\lambda_2,\lambda_2}=0
\end{equation}
\begin{equation}
\mathcal{N}_{\lambda_1,\lambda_1,\lambda_2,\lambda_2}=
\frac{\sqrt{s}}{\Lambda}\beta_t(\lambda_2a_1+\lambda_1a_2\cos\theta)
\end{equation}
\begin{equation}
\mathcal{N}_{0,\lambda_1,-\lambda_2,\lambda_2}=-\lambda_1\lambda_2\frac{a_2}{\Lambda}
\sqrt{\frac{s}{2}}\frac{\gamma_W}{\gamma_t}\beta_W(1+\lambda_1\lambda_2\cos\theta)
\end{equation}
\begin{equation}
\mathcal{N}_{-\lambda_1,0,-\lambda_2,\lambda_2}=\mathcal{N}_{0,\lambda_1,-\lambda_2,\lambda_2}
\end{equation}
\begin{equation}
\mathcal{N}_{0,\lambda_1,\lambda_2,\lambda_2}=\frac{a_2}{\Lambda}
\sqrt{\frac{s}{2}}\gamma_W(\lambda_1\beta_t-\lambda_2\beta_W)\sin\theta
\end{equation}

\begin{equation}
\mathcal{N}_{-\lambda_1,0,\lambda_2,\lambda_2}=\mathcal{N}_{0,\lambda_1,\lambda_2,\lambda_2}
\end{equation}

\begin{equation}
\mathcal{N}_{0,0,-\lambda_2,\lambda_2}=-2a_2\frac{\sqrt{s}}{\Lambda}\frac{\gamma_W^2}{\gamma_t}\beta_W\sin\theta
\end{equation}
\begin{equation}
\mathcal{N}_{0,0,\lambda_2,\lambda_2}=\lambda_2\frac{\sqrt{s}}{\Lambda}\gamma_W^2(a_1(1+\beta_W^2)\beta_t+2a_2\beta_W\cos\theta)
\end{equation}

These expressions completely agree with the ones found
in~\cite{Vincent96}. At high energy the anomalous part is given by
\beqn
\mathcal{N}_{0,0,-\lambda_2,\lambda_2}\simeq
-a_2\frac{m_t}{\Lambda}\frac{s}{M_W^2}\sin\theta, \quad
\mathcal{N}_{0,0,\lambda_2,\lambda_2}\simeq
\lambda_2\frac{\sqrt{s}}{2 \Lambda}
\frac{s}{M_W^2}(a_1+a_2\cos\theta)
\eeqn

This high energy behaviour agrees with the one given
in\cite{LariosYuan}. The angular dependence of $a_2$ classifies
this operator as  $P$-wave, compared to $a_1$ which is
$S$-wave\cite{LariosYuan}.

\end{document}